\documentclass[10pt,epsf,preprint]{aastex}
\input epsf

\newlength{\colwidth}
\newcommand{\lya}{{\mbox Ly$\alpha$~}}
\renewcommand{\H}{\ion{H}{1}}

\newcommand{\Hep}{\ion{He}{2}}

\newcommand{\Cfour}{\ion{C}{4}}
\newcommand{\Sifour}{\ion{Si}{4}}

\lefthead{Bernardi et al..}
\righthead{A feature at $z\sim 3.2$}

\begin{document}


\title{A feature at $z\sim 3.2$ in the evolution of the Ly$\alpha$ forest optical depth}

\author{Mariangela Bernardi\altaffilmark{\ref{CMU},\ref{Chicago}},
Ravi K. Sheth\altaffilmark{\ref{Pitt},\ref{Fermilab}},
Mark Subbarao\altaffilmark{\ref{Chicago}},
Gordon T. Richards\altaffilmark{\ref{PennU}},
Scott Burles\altaffilmark{\ref{MIT}},
Andrew J. Connolly\altaffilmark{\ref{Pitt}},
Joshua Frieman\altaffilmark{\ref{Chicago},\ref{Fermilab}},
Robert Nichol\altaffilmark{\ref{CMU}},
Joop Schaye \altaffilmark{\ref{IAS}}, 
Donald P. Schneider\altaffilmark{\ref{PennU}},
Daniel E. Vanden Berk\altaffilmark{\ref{Fermilab}},
Donald G. York\altaffilmark{\ref{Chicago}},
J. Brinkmann\altaffilmark{\ref{APO}}, and 
Don Q. Lamb\altaffilmark{\ref{Chicago}}
}

\newcounter{address}
\setcounter{address}{1}
\altaffiltext{\theaddress}{\stepcounter{address}
Department of Physics, Carnegie Mellon University, 5000 Forbes Ave., Pittsburgh, PA 15213\label{CMU}}
\altaffiltext{\theaddress}{\stepcounter{address}
University of Chicago, Astronomy \& Astrophysics
Center, 5640 S. Ellis Ave., Chicago, IL 60637\label{Chicago}}
\altaffiltext{\theaddress}{\stepcounter{address}
Department of Physics and Astronomy, University of Pittsburgh, Pittsburgh, PA 15620\label{Pitt}}
\altaffiltext{\theaddress}{\stepcounter{address}
Fermi National Accelerator Laboratory, P.O. Box 500,
Batavia, IL 60510\label{Fermilab}}
\altaffiltext{\theaddress}{\stepcounter{address}
Department of Astronomy and Astrophysics, The Pennsylvania State University, 
University Park, PA 16802\label{PennU}}
\altaffiltext{\theaddress}{\stepcounter{address}
Department of Physics, Massachusetts Institute of Technology, 77 Massachusetts
Avenue, Room 6-113, Cambridge, MA 02139-4307\label{MIT}}
\altaffiltext{\theaddress}{\stepcounter{address}
School of Natural Sciences, Institute for Advanced Study, Einstein Drive, Princeton, NJ 08540\label{IAS}}
\altaffiltext{\theaddress}{\stepcounter{address}
Apache Point Observatory, 2001 Apache Point Road, P.O. Box 59, Sunspot, NM
88349-0059\label{APO}}

\begin{abstract}
The effective optical depth in the Ly$\alpha$ forest region of 1061 
low-resolution QSO spectra drawn from the SDSS database decreases with 
decreasing redshift over the range $2.5\le z\le 4$.  Although the evolution 
is relatively smooth, $\tau_{\rm eff}\propto (1+z)^{3.8\pm 0.2}$,
at $z\sim 3.2$ the effective optical depth decreases suddenly, by about ten 
percent with respect to this smoother evolution.  
It climbs back to the original smooth scaling again by $z\sim 2.9$.  
We describe two techniques, one of which is new, for quantifying this 
evolution which give consistent results.  
A variety of tests show that the feature is not likely to be a consequence 
of how the QSO sample was selected, nor the result of flux calibration or 
other systematic effects.
Other authors have argued that, at this same epoch, the temperature of the 
IGM also shows a departure from an otherwise smooth decrease with time.  
These features in the evolution of the temperature and the optical 
depth are signatures of the reionization of \Hep.
\end{abstract}

\section{Introduction}
\label{Int}

The importance of resonant scattering by neutral hydrogen in the 
intergalactic medium (IGM) was described by Gunn \& Peterson (1965), who 
used the lack of a strong absorption trough in the spectra of high-redshift 
quasars to set limits on the amount of dispersed \H.  Lynds (1971) noted 
that in the spectra of distant quasars there are many absorption features 
blueward of the Ly$\alpha$ emission line; he interpreted the absorption 
features as Ly$\alpha$ lines produced by intervening material. 
The mean absorption in the Ly$\alpha$ forest depends mainly on the gas 
density and the amplitude of the ionising background (Rauch et al. 1997; 
Rauch 1998). The absorption increases rapidly with increasing redshift $z$ 
(e.g., Schneider, Schmidt \& Gunn 1991; Songaila \& Cowie 2002).  

The optical depth $\tau$ for a gas consisting primarily of ionized 
hydrogen and singly ionized helium, which is at density 
$(1+\delta)\equiv \rho/\langle\rho\rangle$ relative to the background 
density $\langle\rho\rangle$ and is in photo-ionization 
equilibrium at redshift $z$, is
\begin{equation}
 \tau(z) \approx 0.7\,\left({\Omega_b h^2\over 0.019}\right)^2\, 
             \left({\Omega_m h^2\over 0.3\times 0.65^2}\right)^{-1/2}\,
             \left({1+z\over 4}\right)^{4.5}\,{T_4^{-0.7}\over \Gamma_{12}}
         \, {(1-Y)\over 0.76}{(1-Y/4)\over 0.94}\,(1+\delta)^2
\label{eq:tau}
\end{equation}
where $\Omega_b h^2$ is the baryon density, 
$H_0=100h$ km s$^{-1}$ Mpc$^{-1}$ is Hubble's constant, 
$\Omega_m$ is the matter density, 
and $Y$ is the helium abundance by mass (e.g. Peebles 1993, \S 23).  
The temperature of the gas is $T_4\equiv T/10^4K$, and 
$\Gamma_{12}\equiv \Gamma/10^{-12}$ s$^{-1}$ is the photo-ionization 
rate.  Equation~(\ref{eq:tau}) suggests that $\tau$ should evolve rapidly.  
Various authors (e.g., Jenkins \& Ostriker 1991; Hernquist et al. 1996; 
Rauch et al. 1997) have noted that measurements of the mean transmission 
$\langle {\rm exp}(-\tau)\rangle$ and its evolution constrain the 
parameters in equation~(\ref{eq:tau}), 
such as the ratio $(\Omega_b h^2)^2/(\Omega_m h^2)^{1/2}$ (Rauch 1998), 
and the evolution of $T_4^{-0.7}/\Gamma_{12}$ 
(McDonald \& Miralda-Escud\'e 2001).  

Equation~(\ref{eq:tau}) shows that, after the reionization of \H, 
the optical depth is expected to decrease smoothly with time, unless, 
for example, there is a sudden injection of energy into the IGM.  
For instance, if the temperature of the gas increases by a factor of two 
at some epoch, then equation~(\ref{eq:tau}) suggests that the optical 
depth $\tau$ in the Ly$\alpha$ forest would decrease by a factor of 
$2^{-0.7}\sim 0.6$.  There is some evidence of a factor of two change in 
the temperature of the IGM at $z\sim 3 - 3.5$ (e.g. Schaye et al. 2001).  

Reimers et al. (1997; also see Heap et al.\ 2000; Kriss et al.\ 2001)
found  evidence for a sharp increase in the \Hep\ opacity around $z\sim 3$, 
which they associated with \Hep\ reionization.  Songaila \& Cowie (1996) 
and Songaila (1998) have argued that the observed evolution of 
\Cfour/\Sifour metal line ratios requires a sudden hardening of the ionizing 
background around $z\sim 3$, which is consistent with \Hep\ reionization.
Schaye et al. (2000) and Theuns et al. (2002a,b) showed that \Hep\ 
reionization at $z\sim 3.5$ results in a jump of about a factor of two 
in the temperature of the IGM at the mean density, and found evidence 
for such a jump by studying the distribution of line-widths in the 
Ly$\alpha$ forest. In addition, Schaye et al. (2000) and 
Ricotti, Gnedin \& Shull (2000) found that the gas is close to isothermal 
at redshift $z\sim 3$, indicating that a second reheating of the 
intergalactic medium took place at $z\sim 3$. This too might be 
interpreted as evidence of the reionization of \Hep.
(Numerical simulations of the observational signatures of \Hep\ ionization 
are also presented in e.g., Meiksin 1994 and Croft et al. 1997.)  
However, Boxenberg (1998) and Kim, Cristiani, \& D'Odorico (2002) 
found no change in \Cfour/\Sifour, and analyses by McDonald et al. (2001) 
and Zaldarriaga, Hui, \& Tegmark (2001) did not find a significant 
temperature change at these redshifts.  
Thus, both from metal line ratios, and from measurements of line widths, 
there is some evidence for \Hep\ reionization at $z\sim 3 - 3.5$, and 
that this event is associated with an increase in the temperature of the 
IGM, although the strength of the evidence is still being questioned.  

If \Hep\ were ionized at $z\sim 3 - 3.5$, and this caused the 
temperature of the IGM to increase by a factor of two, then our simple 
estimate of an associated sixty percent decrease in $\tau$ is not quite 
right. 
For instance, it ignores the fact that the extra electron liberated by the 
ionization can increase the optical depth.  However, for $Y\sim 0.24$, the 
increase in the electron density from the electron released by \Hep\ 
ionization can increase $\tau$ only  by seven or eight percent.  Although 
this goes in the opposite direction to the effect of the temperature 
increase, it is a substantially smaller effect.  Other important factors, 
which the simple sixty-percent estimate ignores, include the facts that 
the temperature change may be accompanied by a change in the 
temperature--density relation of the gas;  
that saturated lines which contribute to the optical depth will not be as 
strongly affected by a temperature change; 
and that a temperature increase may expand the gas, thus affecting peculiar 
velocities and complicating the relationship between temperature, 
line profile and optical depth.  
Nevertheless, the discussion above indicates that a sudden change in 
the temperature of the IGM may well be accompanied by a sudden change 
in the optical depth, although a precise estimate of the magnitude of 
the effect requires hydrodynamical simulations.  

A sudden change in $\tau$ means that the ratio of the mean absorption in 
the Ly$\alpha$ forest to that in the spectrum of the quasar (hereafter 
QSO) should also change abruptly at the same time.  
That is, the quantity defined by Oke \& Korycansky (1982), 
\begin{equation}
 D_A\equiv 1 - \bar F, \qquad {\rm where}\qquad 
 \bar F\equiv {F_\lambda({\rm observed})\over F_\lambda({\rm continuum})}
       \equiv \exp(-\tau_{\rm eff}) 
 \label{eq:DA}
\end{equation}
should show a feature at $z\sim 3 - 3.5$ if \Hep\ was ionized at that time.  
An advantage of studying the mean absorption, $D_A$, or transmission, 
$\bar F$, is that it can be measured even in low resolution spectra for 
which individual line measurements are not possible.  
It is conventional to use the mean transmission to define an effective 
optical depth:  $\tau_{\rm eff}\equiv -\ln \bar F$.  
Schneider, Schmidt \& Gunn (1991) show that the mean transmission evolves 
significantly over the range $0<z<4.5$ (also see 
Press, Rybicki \& Schneider 1993).  
The main goal of this paper is to see if this evolution is smooth, or 
has a feature in it.  For example, we would like to see if there is any 
evidence of a sudden drop in the effective optical depth in the 
Ly$\alpha$ forest at $z\sim 3 - 3.5$.  

Section~\ref{selection} describes how we selected our sample of 
$\sim 10^3$ QSOs from the Sloan Digital Sky Survey (SDSS) database.  
The SDSS QSO selection algorithm itself is studied in some detail in 
Appendix~\ref{seffects}.  We will be searching for a feature in the 
evolution of the mean transmission; this requires an accurate 
determination of the underlying intrinsic QSO spectrum.  This is 
the subject of Section~\ref{contfit}.  We use two methods to do this.  
one method, which is a direct descendent of the one first used by 
Oke \& Korycansky (1982), is described in Appendix~\ref{method2}.  
The other is new; it exploits the fact that the continuum is a function 
of restframe wavelength, whereas the Ly$\alpha$ effective optical depth 
is a function of observed wavelength.  
Both methods yield consistent results---the inferred continuum between 
the Ly$\alpha$ and Ly$\beta$ emission lines is not featureless, and, at 
$z\sim 2.9-3.3$, there appears to be a feature in the otherwise smooth 
evolution of the mean transmission, and hence of $\tau_{\rm eff}$.  
Possible systematic effects which might affect our measurement are 
discussed in Section~\ref{tauz} and in Appendix~\ref{seffects}.  
A final section summarizes our findings.  Appendix~\ref{app:skew} 
is somewhat tangential to the main subject of this paper:  it is a 
short demonstration of some effects which arise from the fact that the 
distribution of flux decrements in the Ly$\alpha$ forest is highly 
non-Gaussian.  

A comparison of this measurement with predictions from hydrodynamical 
simulations shows that our measurements can be interpreted as evidence for 
\Hep\ reionization at $z\sim 3.2$ (Theuns et al. 2002).  
The implications for the evolution of the temperature of the IGM and 
the photo-ionization rate $\Gamma$ will be presented in a future paper.  

\section{Sample selection}\label{selection}

The sample of QSOs used in this paper was extracted from the SDSS 
database (York et al. 2000) which included all the spectra observed 
by the SDSS collaboration through the end of 2001.  This sample is 
about three times larger than that in the SDSS Early Data Release 
(Stoughton et al. 2002).  
The SDSS camera is described in Gunn et al. 1998, and 
the filter response curves are described in Fukugita et al. (1996).  
The SDSS photometric reduction procedure is described in 
Lupton et al. (2000), 
and the spectroscopic data reduction procedure will be described in 
Frieman et al. (2002).
The SDSS procedure for targeting QSOs is described in detail in 
Richards et al. (2002a).  

\begin{figure}[t]
\plotone{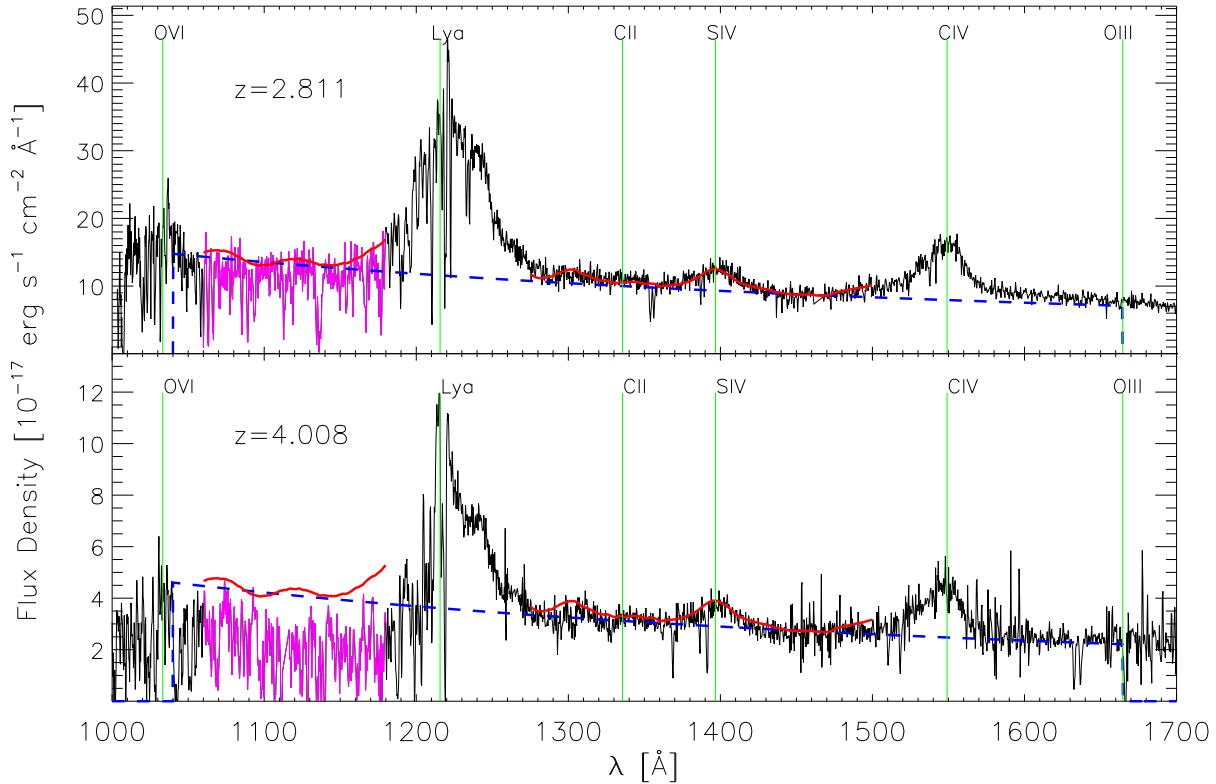}
\caption{Examples of QSO spectra in our sample, shown here as a function 
of wavelength in the restframe. Dashed line shows a power-law in 
wavelength of slope $\alpha_{\lambda}=-1.56$, normalized to have the same 
flux as the observed spectra in the rest wavelength range $1450-1470$~\AA.
Solid line shows the continuum obtained as described in 
Section~\ref{contfit}.  We analyse the Ly$\alpha$ forest in the 
wavelength region 1060--1180~\AA.}
\label{fig:spectra}
\end{figure}

When we selected our sample, the SDSS had imaged $\sim 4000$ square 
degrees, and $\sim 20,000$ QSOs had both photometric and spectroscopic 
information.  
The SDSS spectroscopic pipeline identifies any extragalactic object whose 
spectrum is dominated by a non-stellar continuum and has at least one 
broad emission line (rest-frame FWHM larger than $1000$~kms$^{-1}$) 
as a candidate QSO (Frieman et al. 2002).  Thus, the sample of 
$\sim 20,000$ objects includes Seyfert galaxies and some ``Type 2'' AGNs 
as well as QSOs.  All of these spectra were examined visually to make 
certain that the redshift was correctly assigned.  Spectra with broad 
absorption line features (BAL QSOs) for which it was not possible to 
measure a redshift are not included in the above.  

We are interested in compiling a sample of objects in which the Ly$\alpha$ 
forest can be easily detected. The requirement that the entire forest 
(restframe wavelength $\sim 1050-1180$~\AA) be detected in the SDSS 
spectrum of an object sets a lower limit of $z\approx 2.75$ on the 
redshift of the QSOs we will analyze.  In turn, this sets a limit of 
about $z_\alpha\ge 2.5$ on the redshift range in which we can study 
the Ly-$\alpha$ forest.  In practice, there is a problem with the flux 
calibration of the SDSS spectra at the blue end of the spectrograph 
(wavelengths shorter than 4400\AA, see Appendix~\ref{app:Civ}.  
Therefore, we only show results at slightly higher redshifts, which are 
not affected by this.  
Of the $\sim 20,000$ QSO--candidate objects above, about $1400$ are at 
$z\ge 2.75$.  About 250 of these had spectra with unusually broad absorption 
lines (BALs), or strong damped Ly$\alpha$ systems, and/or had low quality 
spectra, so we removed them from our sample.  

The instrumental resolution of the SDSS spectrograph is about 
$150$~kms$^{-1}$.  Studies of higher resolution QSO spectra show that 
most lines in the Ly$\alpha$ forest are substantially narrower than this.  
Therefore, a typical line in the Ly$\alpha$ forest is unresolved in our 
data.  In addition, the SDSS QSO spectra have a median $S/N$ per pixel 
of $\sim 10$, and this ratio drops to $\sim 3$ in the Ly$\alpha$ forest.  
Therefore, for the SDSS sample, measuring the parameters of individual 
Ly$\alpha$ lines as a function of redshift is not the best way to estimate 
the temperature evolution of the IGM.  
A better approach is to measure the mean transmission of the flux 
in the Ly$\alpha$ forest, $\bar F$, as a function of redshift.  
Equation~(\ref{eq:DA}) shows that the crucial step is to determine the 
QSO continuum precisely.  To do so, we must have a reasonably long 
restframe wavelength range which is common to all the objects in our 
sample---we require that the restframe range $1250-1665$~\AA\ be 
detected in all the spectra we will include in our sample.  
This sets an upper limit $z\approx 4.3$ on the redshifts of 
the objects we will include in our analysis.  
This requirement removed an additional $\sim 100$ objects, leaving 
1061 QSO spectra in our sample; two examples are shown in 
Figure~\ref{fig:spectra}.  

\section{Estimating the continuum and the mean transmission}\label{contfit}
In what follows, it will be useful to think of the observed flux in the 
spectrum of the $i$th QSO (shifted to the restframe of 
the QSO and normalized in some standard fashion which we 
will discuss shortly) as  
\begin{displaymath}
 f_i(\lambda_{rest}) = 
   \Bigl[C(\lambda_{rest}|z_i) + c_i(\lambda_{rest})\Bigr]\,
          \Bigl[T(z_\alpha) + \ t_i(z_\alpha)\Bigr] 
           + n_i(\lambda_{obs}),\qquad {\rm where}\qquad 
          1+z_\alpha \equiv {\lambda_{obs}\over \lambda_\alpha}
                     = {\lambda_{rest}(1+z_i)\over 1215.67},
\end{displaymath}
$z_i$ is the redshift of the QSO and $\lambda_\alpha\equiv 1215.67$~\AA.  
Here $C$ represents the mean continuum at fixed rest wavelength, 
which we think of as being representative of the QSO population at $z_i$ 
as a whole (if QSOs evolve, then the mean continuum of the population may 
depend on redshift), and $c_i$ represents the fact that the continuum of 
the $i$th QSO might be different from the mean at that redshift.  ($c_i$ 
could also differ from one QSO to another if relativistic outflows from 
QSOs are common.  See Richards et al. 1999 and references therein 
for evidence of such relativistic velocities.)  
Similarly, $T(z_\alpha)\equiv \exp[-\tau_{\rm eff}(z_\alpha)]$ is the 
mean transmission through the Ly$\alpha$ forest at $z_\alpha$, averaged 
over all the $z_\alpha$ pixels in the forest (note that $\tau_{\rm eff}$ 
is a function of $z_\alpha$, and hence of the observed rather than 
restframe wavelength), and $t_i$ represents the fact that the transmission 
through the forest along the $i$th line of sight might be different from 
the mean value.  The final term $n_i$ represents the noise in the 
observation.  By definition $\langle c_i\rangle\equiv 0$ and 
$\langle t_i\rangle\equiv 0$, where the average over $c_i$ is over fixed 
$\lambda_{rest}$, and the average of $t_i$ is over fixed $z_\alpha$, and 
hence over fixed $\lambda_{obs}$.  We will assume that, at fixed 
$\lambda_{obs}$, $\langle n_i\rangle = 0$ also.  
We have written the observed flux in this way to emphasize the fact that 
the mean continuum $C$ is a function of $\lambda_{rest}$, whereas the mean 
transmission $T=\exp(-\tau_{\rm eff})$ is a function of $\lambda_{obs}$.  
It is this fact which makes it possible, at least in principle, to 
disentangle the two unknown functions $C$ and $T$ from the single observed 
quantity, $f$.  

All work to date first estimates $C + c_i$, and then averages all the 
$f_i/(C+c_i)$ which have the same $z_\alpha$ to estimate the mean 
transmission.  That is, the shape of the continuum is determined separately 
for each QSO.  This is easier to do at low redshifts $z\le 1$ where 
absorption by the forest is smaller, but it is considerably more difficult 
at higher redshifts.  Furthermore, if the resolution of the spectrograph 
is low and/or the signal-to-noise ratio is poor, then systematic errors in 
the estimated continuum can arise (Steidel \& Sargent 1987).  Biases can 
also arise if some fraction of the absorption is not due to \H\ but to 
other elements.  This extra absorption becomes increasingly important at 
lower redshifts, as the \lya opacity decreases more rapidly than the 
opacity of the metals.  For example, at $z\sim 2.5$, approximately 20\% 
of the total absorption in the \lya forest is not due to \H\ 
(Kulkarni et al. 1996; Rauch 1998).  Our sample is confined to high enough 
redshifts that this should not be a significant concern, although, as we 
discuss later, absorption by elements other than \H\ may be important 
when comparing our measurements to results from higher resolution spectra.  

Our spectra have low resolution and signal-to-noise, so an object-by-object 
estimate of the continuum is difficult.  On the other hand, our sample is 
very large, so we can take a statistical approach.  
Consider QSOs in a small redshift range.  The QSOs have a range of 
luminosities, and so the set of spectra in any one redshift bin can differ 
considerably from each other.  When suitably normalized, however, the 
differences between spectra are reduced significantly.  Therefore, 
following Press, Rybicki \& Schneider (1993) and Zheng et al. (1997), we 
normalize each spectrum by the flux in the rest wavelength range 
$1450-1470$~\AA.  (This wavelength range lies in front of the 
\ion{C}{4} emission line, and is free of obvious emission and absorption 
lines.)  
Having normalized each observed spectrum we compute the average value of 
the normalized flux $f_i$ to obtain a composite spectrum.  
This composite is 
\begin{displaymath}
 \langle f_i\rangle = CT + \langle Ct_i\rangle + \langle c_iT\rangle + 
                           \langle c_it_i\rangle + \langle n_i\rangle = CT,
\end{displaymath}
where we have assumed that the averages $\langle Ct_i\rangle$, 
$\langle c_iT\rangle$, $\langle c_it_i\rangle$ and $\langle n_i\rangle$, 
evaluated at fixed $\lambda_{rest}$ and QSO redshift are all zero.  For 
a sufficiently large sample, these averages probably are vanishingly 
small, so we can interpret the measured composite spectrum as the product 
of the mean continuum times the desired mean transmission.  Because 
 $\langle f_i/C\rangle = \langle f_i/(C+c_i)(1+c_i/C)\rangle = 
  \langle f_i/(C+c_i)\rangle + \langle f_i/(C+c_i)(c_i/C)\rangle$, 
this estimate of the mean transmission differs from the usual one by 
the second term:  $\langle f_i/(C+c_i)(c_i/C)\rangle$.  
If the transmission $f_i/(C+c_i)$ in the $i$th spectrum is not correlated 
with how different the continuum of the $i$th QSO is compared to the 
average continuum, $c_i/C$, then this second term can be written as two 
separate averages.  In this case, 
$\langle f_i/C\rangle = \langle f_i/(C+c_i)\rangle$ because 
$\langle c_i/C\rangle \equiv 0$.  

\begin{figure}[t]
\plotone{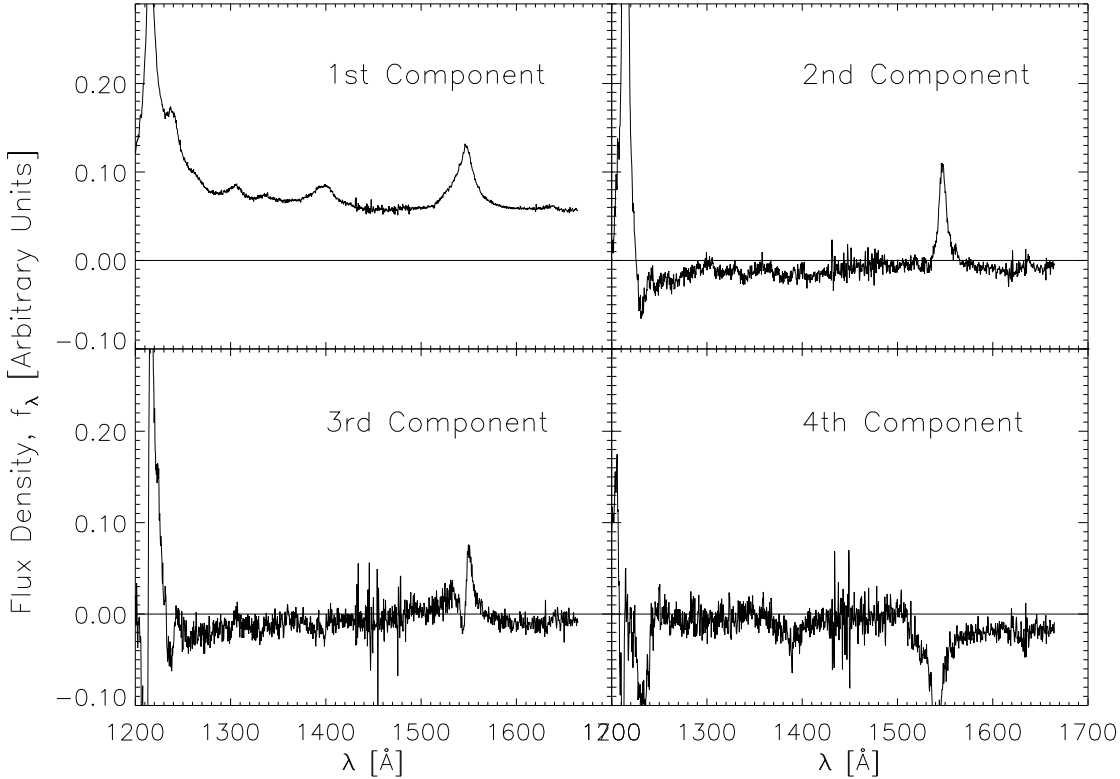}
\vspace{-5.5cm}
\caption{First four components (or eigen-spectra) obtained from 
a principle component analysis (PCA) of all the QSO spectra in 
the redshift bin $2.9 < z < 3.3$.  Eigen-spectra for the other 
redshift bins are similar.  The second and higher order components 
mainly try to fit the emission line features accurately; the overall 
shape is determined primarily by the first component.}
\label{fig:pca}
\end{figure}

If QSOs at the same redshift have a wide variety of spectra, then the 
composite spectrum could be very different from the spectrum of any 
individual object, thus making our estimates of the mean transmission 
blueward of $\lambda_\alpha$ very noisy.  
Therefore, we carried out a principle component analysis (PCA; e.g., 
Francis et al. 1992) of the spectra in the wavelength range 
$1200-1665$~\AA.  Figure~\ref{fig:pca} shows the first four components 
(or eigen-spectra) determined by the PCA for the QSOs in the redshift 
range $2.9 < z < 3.3$ (the other redshift bins show similar eigen-spectra). 
The first component (upper left panel) represents the intervals 
$1280-1500$~\AA\ and $1580-1665$~\AA\ well.  The next $10-15$ components 
are mainly necessary for reproducing the exact shape of the Ly$\alpha$ 
and \ion{C}{4} emission lines.  In particular, because the flux density 
of these higher order components is close to zero in the 
$1280-1500$~\AA\ and $1580-1665$~\AA\ wavelength regions, the PCA 
analysis suggests that, in these regions QSO spectra are very similar to 
each other.  
Therefore, our decision to combine all the QSOs at a given redshift when 
estimating the shape of the continuum (i.e., to treat all QSOs at 
a given redshift as differing in the normalization, but not the shape, 
of the continuum) is likely to be reasonable.  

\begin{figure}
\plotone{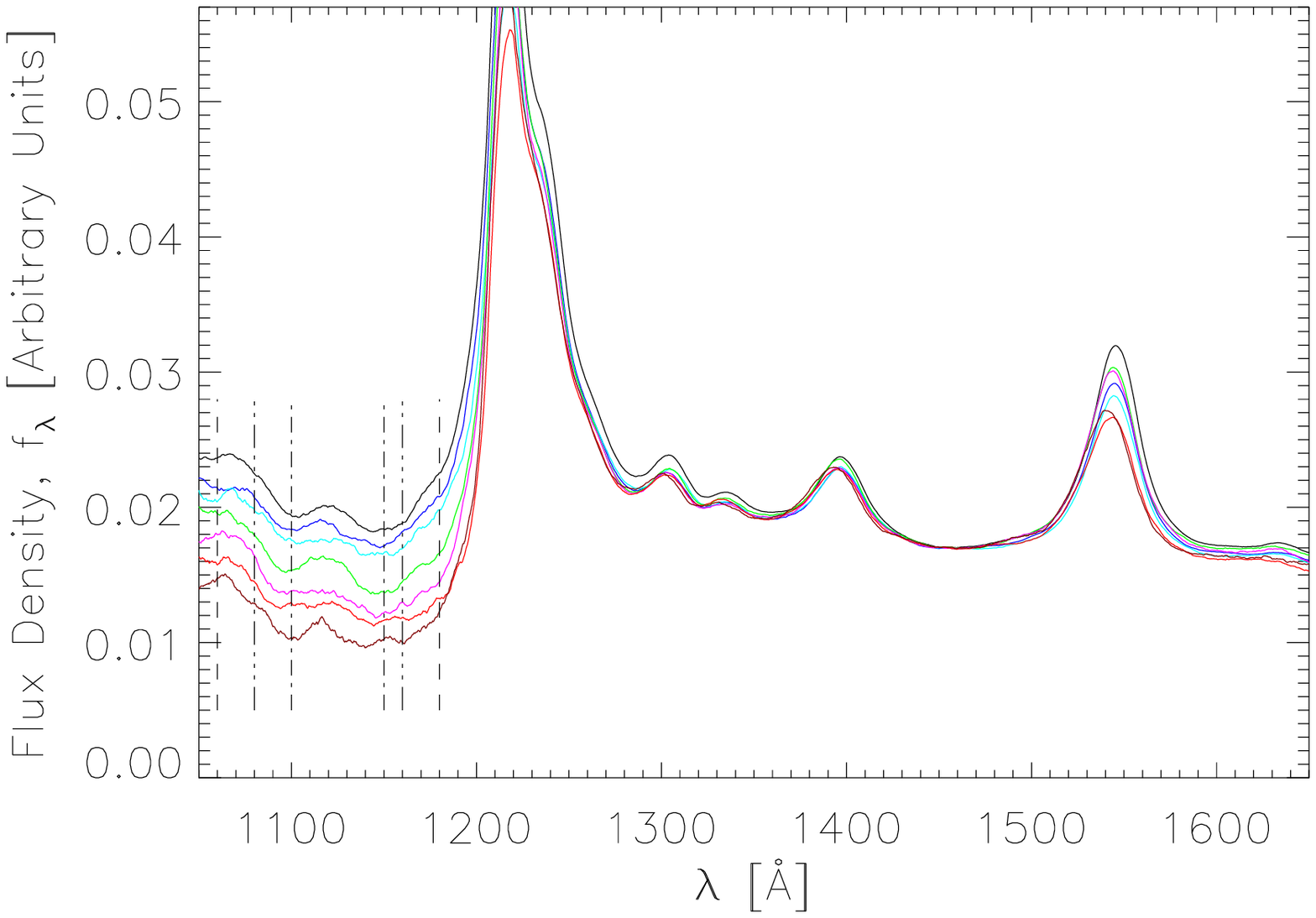}
\caption{Composite restframe spectra (i.e. wavelengths have been 
transformed to the restframe of the QSO, and flux densities were normalized 
to sum to the same value in the range $1450-1470$~\AA) in different 
redshift bins; curve which is highest at small $\lambda$ is for 
$2.95\le z < 3.15$, and bins in redshift step by $\Delta z=0.2$ down 
to the lowest curve which is for $4.15\le z < 4.3$.  Redward of the 
Ly$\alpha$ emission line, the different composites are extremely similar.  
In contrast, the region blueward of $\lambda_\alpha = 1215.67$~\AA\ 
changes rapidly with redshift.  The bumps at $\lambda = 1070$~\AA\ and 
$\lambda=1120$~\AA\ are emission features intrinsic to the QSO spectrum.
The vertical lines on the left show three different 
wavelength regions adopted in defining the Ly$\alpha$ forest: 
1060--1180~\AA\ (dashed), 1080-1160~\AA\ (dot-dot-dot-dashed), 
and 1100--1150~\AA\ (dot-dashed).}
\label{compz}
\end{figure}

Figure~\ref{compz} shows composite spectra as a function of restframe 
wavelength for a number of bins in redshift:  the curve which is highest 
on the left is for $2.95\le z < 3.15$, the next highest is for 
$3.15 \le z < 3.35$, and so on, until the lowest curve which is 
for $4.15\le z< 4.3$.  The bins in $\lambda$ over which this averaging 
was done were chosen to be as small as possible---they were set by the 
SDSS pixel sizes.  
There are typically about 100 QSOs per bin (Figure~\ref{fig:NzQSO} shows 
the exact distribution of QSO redshifts in our sample).  
The vertical lines on the left show three different 
wavelength regions adopted in defining the Ly$\alpha$ forest: 
1060--1180~\AA\ (dashed), 1080-1160~\AA\ (dot-dot-dot-dashed), 
and 1100--1150~\AA\ (dot-dashed).

Redward of the Ly$\alpha$ emission line at $\lambda_\alpha=1215.67$~\AA, 
the different curves in Figure~\ref{compz} are all very similar to each 
other (although the \ion{C}{4} emission line and redward may be evolving 
slightly, and the uppermost curve, corresponding to the QSOs in the lowest 
redshift bin, appears to be slightly different from all the others; the 
apparent evolution of the red wing of the \ion{C}{4} line is discussed 
in more detail by Richards et al. 2002b).  
Evidently, redward of $\lambda_\alpha$, the QSO population as 
a whole evolves little between $z\sim 4.3$ and $z\sim 3$.   Blueward of 
$\lambda_\alpha$, however, there is an obvious trend:  there is less 
observed flux in the spectra of higher redshift QSOs.  Our problem is 
to turn this trend into a quantitative estimate of how the effective 
optical depth evolves.  This can be done because the effective optical 
depth is the same at fixed observed, rather than restframe wavelength, 
whereas the continuum is a function of restframe wavelength.  

To illustrate, consider the bumps at 1070~\AA\ and 1120~\AA.  Because 
they are present at the same restframe wavelengths in all the redshift 
bins, they cannot have been caused by features in the evolution of the 
optical depth.  (If we define the Ly$\alpha$ forest as spanning the range 
$1060-1180$~\AA, then for the lowest redshift bin, $2.95\le z < 3.15$, 
the forest spans the range $2.44\le z_\alpha< 3.03$, whereas 
for the highest redshift bin, $4.15\le z< 4.3$, the forest spans 
$3.49\le z_\alpha< 4.14$.  Features at fixed $z_\alpha$ would appear at 
quite different wavelengths for the different QSO redshift bins.)  
Therefore, the bumps must be intrinsic to the QSO spectrum---they 
could be \ion{Ar}{1} and \ion{Fe}{3} in emission.  Their presence 
can affect our estimates of the effective optical depth in the forest.  

In low resolution observations such as ours, the continuum level is 
usually calibrated redwards of the \lya emission line and then 
extrapolated bluewards assuming a smooth power-law shape (e.g., 
Press, Rybicki \& Schneider 1993).  However, departures from a smooth 
power-law, such as the two emission lines at $\sim 1070$~\AA\ and 
$\sim 1120$~\AA, are clearly present in our data.  
(Figure~5 of Press, Rybicki \& Schneider 1993 also shows bumps at these 
wavelengths, although they do not call attention to them.)  
In addition, at wavelengths close to the Ly$\alpha$ or Ly$\beta$/\ion{O}{6} 
emission features, emission from the QSO can contaminate the flux in the 
forest---smooth power-law fits to the mean continuum shape cannot account 
for this.  The following section describes how we solve simultaneously 
for the evolution of the effective optical depth and for the shape of the 
mean continuum, while allowing for the possibility that neither are 
well-fit by featureless power-laws.  

\subsection{Method:  A minimization approach}\label{secchi2}
The observed composite spectrum is the product of the mean continuum 
times the mean transmission.  This fact suggests defining 
\begin{equation}
 \chi^2 = \sum_{i} \left(f_i(\lambda_{rest}) -
    C(\lambda_{rest})\,
    {\rm e}^{-\tau_{\rm eff}(\lambda_{obs}/\lambda_\alpha)}\,\right)^2,
\label{chisq}
\end{equation}
where the sum is over all pixels in all spectra in the sample which fall 
in the wavelength range associated with the forest.  The composite spectra 
suggest that the continuum is the superposition of a power-law, two emission 
lines and the blueward side of the Ly$\alpha$ emission line.  
Therefore, we parametrize 
\begin{displaymath}
 C = c_0\,\left(\lambda_{rest}\over \lambda_\alpha\right)^{c_1} 
       + c_2\,\exp\left(-{(\lambda_{rest}-c_3)^2\over 2\,c_4^2}\right)
       + c_5\,\exp\left(-{(\lambda_{rest}-c_6)^2\over 2\,c_7^2}\right)
       + c_8\,\exp\left(-{(\lambda_{rest}-c_9)^2\over 2\,c_{10}^2}\right), 
\end{displaymath}
and we set
\begin{displaymath}
 T=\exp(-\tau_{\rm eff})\qquad {\rm with}\qquad
 \tau_{\rm eff} = 
 t_0\,\left({\lambda_{rest}(1+z_i)\over \lambda_\alpha}\right)^{t_1} + 
 t_2\,\exp\left(-{(\lambda_{rest}(1+z_i)/\lambda_\alpha - t_3)^2\over 2\,t_4^2}\right),
\end{displaymath}
so as to allow the possibility of a feature centred at $1+z_\alpha=t_3$ 
superimposed on an otherwise smooth power-law evolution of $\tau_{\rm eff}$.  
To reduce the number of free parameters we have fixed the position of
the peak of the Ly$\alpha$ emission line ($c_9 = 1215.67$~\AA) and of the 
other two emission lines seen in the composite spectrum ($c_3 = 1073$~\AA\
and $c_6 = 1123$~\AA).
Then the remaining eight parameters of $C$ and the five parameters of 
$\tau$ are varied until $\chi^2$ has been minimized (note that 
$\lambda_\alpha=1215.67$ is not a parameter).  
In principle, we could have attempted to fit the 
emission lines redward of $\lambda_\alpha$ (as was done by 
Press, Rybicki \& Schneider 1993).  Since we are more interested in the 
shape of the continuum blueward of $\lambda_\alpha$, we did not do this.  

The exact parameter values which minimize $\chi^2$ depend somewhat on the 
range used to define the Ly$\alpha$ forest.  We have tried three ranges 
which are shown in Figure~\ref{compz}:  the largest range 
1060--1180~\AA\ (shown by the dashed lines) requires that we understand 
the continuum even in the regime which is close to the Ly$\alpha$ emission 
line, the shorter less demanding range 1080--1160~\AA\ (dashed-dot-dot-dotted 
lines) is our standard, and the shortest, most conservative range is 
1100--1150~\AA\ (dashed-dotted lines).  

\begin{table}[t]
\begin{centering}
\label{tab1}
\caption{Values which minize $\chi^2$ as defined by equation~(\ref{chisq}).  
Two sets of values are shown: first is for the entire sample, second is for 
a subset which has a higher signal-to-noise ratio.}
\begin{tabular}{lccccccccccccc}
 & & & & & & & & & & & & \\ 
\tableline 
\tableline
 All & & $c_0$ & $c_1$ & $c_2$ & $c_3$ & $c_4$ & $c_5$ & $c_6$ & $c_7$ & $c_8$ & $c_9$ & $c_{10}$ \\ 
\tableline 
 1061& &0.0220 & $-1.56$ & 0.0023 & 1073 & 11 & 0.0022 & 1123 & 13 & 0.025 & 1216 & 25 \\
 & $\pm $ & $ 0.0006$ & $ \ \ 0.02$ & $ 0.0003$ & -- & $ 1$ & $ 0.0002$ & -- & $ 2 $ & $ 0.008$ & -- & $ 1$ \\  
\tableline
 &  & & & $t_0$ & $t_1$ & $t_2$ & $t_3$ & $t_4$ & & & \\ 
\tableline 
  &  & & & 0.0028 & 3.69 & $-0.06$ & 4.15 & 0.08 & & & \\
 & & & $\pm$ & $ 0.0011$ & $  0.22$ & $ \ \ 0.02$ & $ 0.02$ & $ 0.02$ & & & \\
\tableline
\tableline
 $S/N>4$ &  & $c_0$ & $c_1$ & $c_2$ & $c_3$ & $c_4$ & $c_5$ & $c_6$ & $c_7$ & $c_8$ & $c_9$ & $c_{10}$ \\ 
\tableline 
 796 & & 0.0224 & $-1.56$ & 0.0033 & 1073 &  9 & 0.0023 & 1123 & 9 & 0.021 & 1216 & 29 \\
 & $\pm$ & $  0.0008$ & $ \ \ 0.02$ & $ 0.0003$ & -- & $ 1$ & $ 0.0002$ & -- & $ 2 $ & $ 0.006$ & -- & $ 2$ \\  
\tableline
 &  & & & $t_0$ & $t_1$ & $t_2$ & $t_3$ & $t_4$ & & & \\ 
\tableline 
 &  & & & 0.0024 & 3.79 & $-0.09$ & 4.14 & 0.09 & & & \\
 & & &$\pm$  & $ 0.0014$ & $ 0.18$ & $ \ \ 0.02$ & $ 0.03$ & $ 0.02$ & & & \\
\tableline
 \end{tabular}
 \end{centering}
\end{table}

\begin{figure}
\plotone{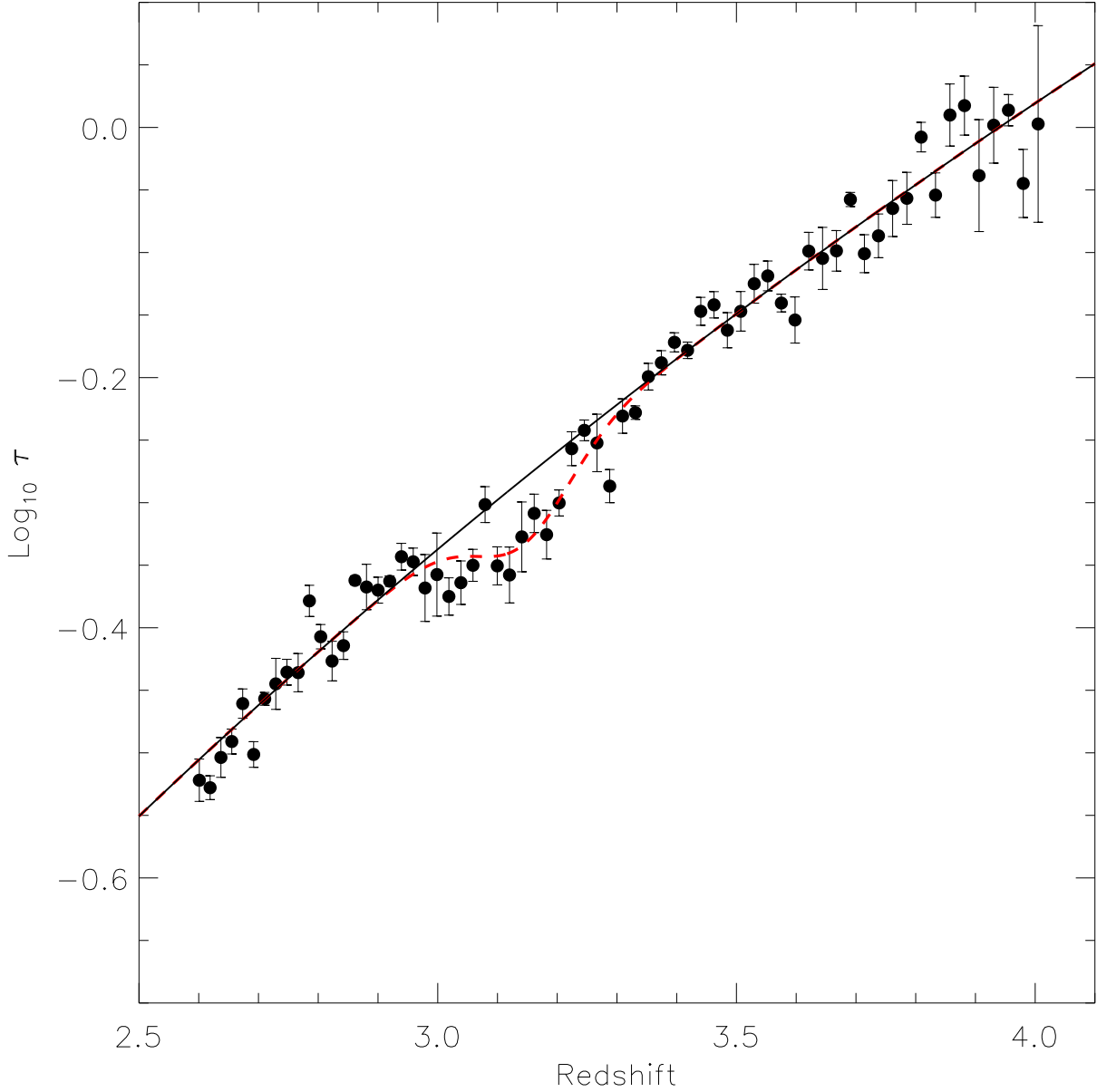}
\caption{The effective optical depth as a function of redshift 
$z_\alpha = \lambda_{obs}/\lambda_\alpha - 1$:  
the average is over all pixels which have the same $z_\alpha$ and 
which have restframe wavelengths which lie in the Ly$\alpha$ forest 
region between $1080-1160$~\AA.  
Dashed line shows the mean evolution obtained from the $\chi^2$ 
procedure described in Section~\ref{secchi2}.  
Filled circles show the estimate from the iterative technique described 
in Appendix~\ref{method2}.  
Solid line shows a simple power-law:  
$\tau_{\rm eff}\propto (1+z_\alpha)^{3.69}$.}
\label{meantau}
\end{figure}

The parameters which minimize $\chi^2$ for our standard definition of the 
wavelength range spanned by the forest ($1080-1160$~\AA) are given in 
Table~1.  (Two technical comments are necessary.  
First, only the longest wavelength range has sufficient wavelength 
coverage to constrain well all three Gaussians which make up the continuum.  
Therefore, in practice, the parameters which define the continuum were 
set using the largest wavelength range, $1060-1180$~\AA, and were held 
fixed when analyzing the standard and the shorter ranges.  Second, as we 
show in Appendix~\ref{app:Civ} below, there is a calibration problem at 
the very blue end of the spectrograph.  This affects the pixels 
corresponding to the redshifts $z_\alpha< 2.5$ in the forest, so the 
minimization procedure was run using only pixels with $z_\alpha\ge 2.6$.)  
Two sets of values are shown: the first is for the entire sample, and 
the second is for a subset which has a higher signal-to-noise ratio 
(see Section~\ref{snratio}).  
The quoted errors are from bootstrap resampling, with replacement, of 
entire QSO spectra.  
The slope of the continuum, $c_1=-1.56$ is the same as that reported 
by Vanden Berk et al. (2001) in their analysis of SDSS QSOs.  
The exact shape is shown by dotted lines in Figure~\ref{fig:iter}.
The smooth evolution of the optical depth, $t_0=0.0028$ and $t_1=3.69$, 
is in reasonable agreement with previous analyses of low resolution 
spectra (e.g. Press, Rybicki \& Schneider 1993).  The fact that 
$t_2=-0.060$ is less than zero suggests that $\tau_{\rm eff}$ decreases 
by about 10 percent at $z_\alpha=t_3-1=3.15$.  
Figure~\ref{meantau} shows this feature superimposed on the otherwise 
smooth evolution of the effective optical depth.  

The procedure above requires minimization of a function which depends 
nonlinearly on the parameters to be fitted.  
Press, Rybicki \& Schneider (1993) discuss how and why one might 
approximate a nonlinear function of the sort above by one which depends 
linearly on the parameters to be fitted.  Since we have a good idea of 
where the features in the continuum and in the mean transmission might 
be (from the composite spectra), we could experiment with performing 
simpler linear fits of the sort they discuss, although we have not done 
so here.  
A modification to the method, which we have also not explored, is to 
weigh each pixel by the inverse of the noise when defining $\chi^2$.  

This method is very different from any in the literature.  
In Appendix~\ref{method2}, we describe a technique which is more closely 
related to that introduced by Oke \& Korycansky (1982), and developed 
further by Schneider, Schmidt \& Gunn (1991) and 
Press, Rybicki \& Schneider (1993).  The next subsection shows that 
both techniques give consistent results.  

\subsection{Result: The evolution of the effective optical depth}\label{meant}
Having determined how the mean continuum depends on restframe wavelength, 
and how the mean transmission depends on redshift $z_\alpha$, we can 
now estimate how the effective optical depth 
$\tau_{\rm eff}\equiv -\ln \bar F$ evolves (note that we used $T$ for 
$\bar F$ in the previous subsections).  
The dashed line in Figure~\ref{meantau} shows the effective optical 
depth obtained from the $\chi^2$ technique described in the previous 
section (i.e., from the parameter values in Table~\ref{tab1}).  
To highlight the feature at $z\sim 3.2$, the solid line shows the 
smoother function 
$\tau_{\rm eff}\propto (1+z_\alpha)^{3.69}$.

The solid circles in Figure~\ref{meantau} show $\tau_{\rm eff}$ as a 
function of $z_\alpha$, estimated using the iterative technique 
described in Appendix~\ref{method2}.  The different circles show 
averages over all pixels which have the same observed wavelength 
$\lambda_{obs}=\lambda_\alpha(1+z_\alpha)$, and which have restframe 
wavelengths which lie in the Ly$\alpha$ forest region between 
$1080-1160$~\AA.  The error bars were computed by bootstrap re-sampling, 
with replacement, the entire sample 50 times (entire QSO spectra, 
rather than individual pixels, are re-sampled).  The error bars show 
the standard deviation of the 50 mean values.  The bootstrap procedure 
also allows an estimate of bin-to-bin correlations:  each of the circles 
in Figure~\ref{meantau} is correlated with its first nearest neighbour 
on either side, but the covariances fall rapidly for more distant pairs.  

Figure~\ref{meantau} shows that our estimates of the evolution of 
the effective optical depth are in good agreement with each other 
(compare dashed line with solid circles).  Although 
$\tau_{\rm eff}(z_\alpha)$ increases with increasing redshift, 
there is a statistically significant change in the evolution 
around $z_\alpha\approx 2.9-3.2$.  
Flux calibration problems are not the origin of this feature (see 
Appendix~\ref{app:Civ}). Although our two techiniques might produce 
small systematic errors in the determination of the mean transmission, 
these errors are not expected to produce such a relatively sudden 
change as a function of redshift.  

\begin{figure}
\plotone{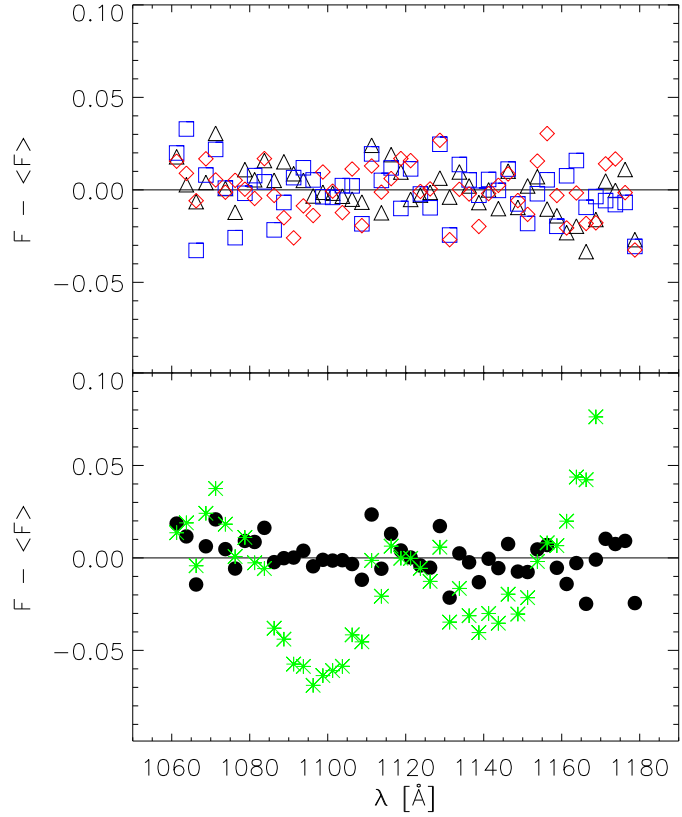}
\vspace{-5.5cm}
\caption{Difference from the mean transmission at a given redshift 
$z_\alpha$, plotted as a function of restframe wavelength rather than 
redshift.  Triangles, squares and diamonds in the top panel 
show measurements averaged over QSOs 
at low ($z<3.2$), medium ($3.2<z<3.7$) and high ($z>3.7$) redshifts. 
Top panel shows no trends, suggesting that our estimate of the 
continuum is reasonable.  Filled circles in bottom panel show the 
average over all redshift bins, and stars show the same test but using 
the featureless $\alpha_\lambda = -1.56$ power-law continuum.  In this 
last case there is significant structure, illustrating that neglecting 
the emission features is a bad approximation.  }
\label{prox}
\end{figure}

To test if our estimate of the continuum shape is reasonable, we computed 
the residual of each pixel at $z_\alpha$ from the mean transmission at 
$z_\alpha$.  If we have estimated the continuum (and hence the mean 
transmission $T$) correctly, then a plot of the residuals $f_i/C - T$ 
versus $\lambda_{rest}$ (rather than $\lambda_{obs}$, which is 
effectively what the x-axis in Figure~\ref{meantau} is) should not show 
any trend.  (Recall that 
$\langle f_i/C - T\rangle = \langle f_i/(C+c_i) + f_i/(C+c_i)(c_i/C)\rangle - T
         = \langle t_i\rangle -T + \langle f_i/(C+c_i) (c_i/C)\rangle 
         = \langle f_i/(C+c_i)\rangle \langle c_i/C\rangle = 0$, 
and that the $\chi^2$ method is constructed to satisfy this condition.)
Triangles, squares and diamonds in the top panel in Figure~\ref{prox} show 
the mean value of the measurement averaged over the pixels from QSOs at 
low ($z<3.2$), medium ($3.2<z<3.7$) and high redshift ($z>3.7$).  
The absence of any trends suggests that our estimate of the continuum 
is, indeed, accurate.  Just for comparison, the stars in the bottom panel 
show the mean of the residuals computed using the featureless 
$\alpha_\lambda=-1.56$ power-law continuum.  Note the structures which 
coincide with the positions of the emission lines discussed previously.  

In summary:  we have described two methods which allow one to solve 
simultaneously for the shape of the QSO continuum in the restframe 
wavelength range between the Ly$\alpha$ and Ly$\beta$ emission lines 
and the evolution of the mean transmission in the Ly$\alpha$ forest.  
The two methods lead to the same conclusion---the inferred continuum is 
not a featureless power-law but has bumps in it; these are almost certainly 
emission lines from the QSO.  It is important to account for these 
features in the continuum when estimating the mean transmission in the 
Ly$\alpha$ forest.  Although the effective optical depth increases with 
increasing redshift, it does not evolve smoothly:  there appears to 
be little or no evolution around $z_\alpha\sim 3$.  The next section 
studies the evidence for this feature in more detail.  

\section{Tests of systematic effects on the estimated evolution}\label{tauz}
This section discusses a number of possible systematic effects which 
might have given rise to a feature in the mean transmission, but 
argues that none of these are the cause.  

\subsection{The QSO selection algorithm}
\begin{figure}
\plotone{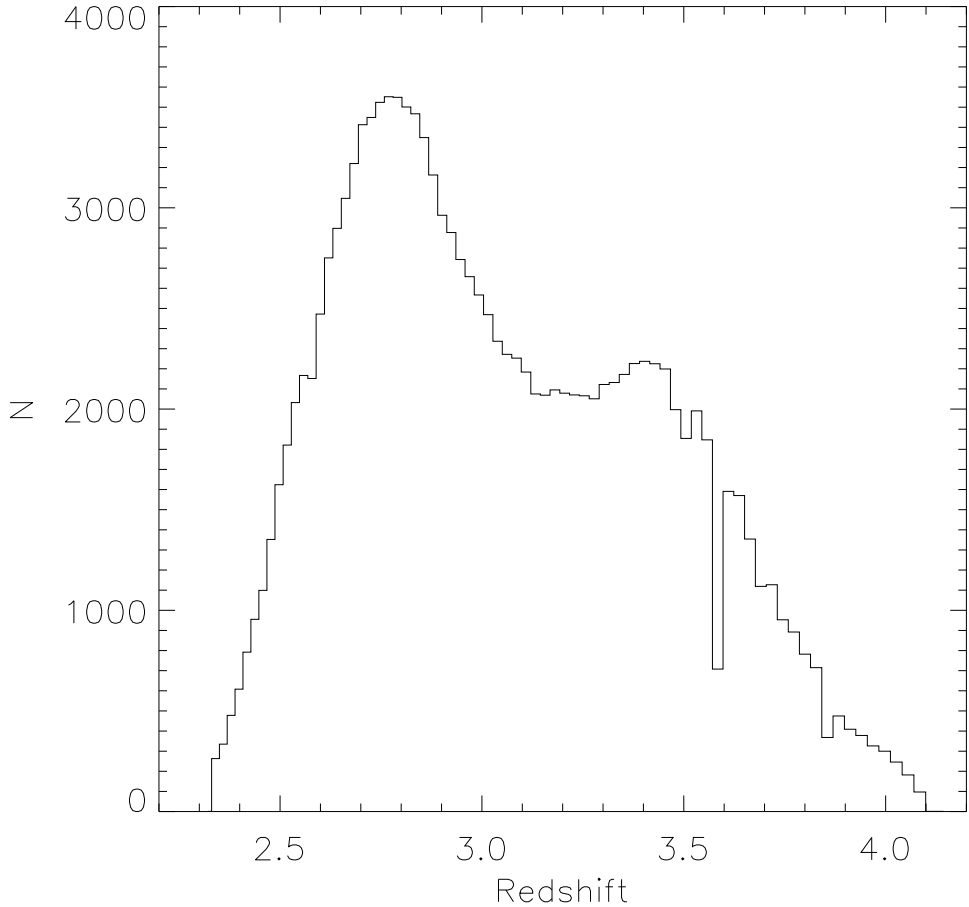}
\vspace{-3.cm}
\caption{Distribution of Ly$\alpha$ forest redshifts $z_\alpha$ in our 
sample.  Each pixel at observed wavelength $\lambda_{obs}$, which has 
a restframe wavelength in the Ly$\alpha$ forest region defined in the 
main text, is assigned a redshift 
 $z_\alpha = \lambda_{obs}/\lambda_\alpha - 1$.  There are 
typically of order 120 pixels per spectrum which lie in the Ly$\alpha$ 
forest; with $\sim 1000$ spectra, this means there are about 120,000 
Ly$\alpha$ forest pixels in total.  The drop in numbers around $z\sim 3.2$ 
is a consequence of the  SDSS QSO selection procedure, as discussed in 
Appendix~\ref{seffects}.  The gaps at $z\sim 3.59$ and $z\sim 3.84$ 
correspond to the \ion{O}{1}~(5577~\AA) sky-line and interstellar 
\ion{Na}{1}~(5894.6~\AA), respectively; 
the observed wavelength range $5570 \le \lambda \le 5590$~\AA\ was removed 
from our analysis, as were the pixels affected by the \ion{Na}{1} line.  }  
\label{fig:hist}
\end{figure}

The $\chi^2$ method solves simultaneously for the mean transmission and 
the mean continuum.  The technique works best when the sample covers a 
large range in redshift. Figure~\ref{fig:hist} shows the number 
of pixels in our sample at each redshift $z_\alpha$, when the forest is 
defined by our standard wavelength interval 1080--1160~\AA\ (the middle of 
the three intervals shown in Figure~\ref{compz}).  
Each spectrum contains about 120 pixels which fall in the Ly$\alpha$ 
forest, and we have on the order of $10^3$ spectra.  Therefore, we have 
a large number of pixels from which to determine the shape of the 
continuum and the transmission, and the figure shows that they do indeed 
span a large redshift range.  

However, two features in Figure~\ref{fig:hist} deserve further comment.  
First, there are obvious drops at $z\sim 3.59$ and at $z\sim 3.84$.  
The SDSS pipeline reductions do not completely subtract the sky-line 
\ion{O}{1}~($5577$~\AA).  This wavelength range corresponds to a Ly$\alpha$ 
redshift of $z\sim 3.59$; hence the gap in the Figure.  Therefore, we 
removed from our analysis the observed wavelength range 
$5570 \le \lambda \le 5590$~\AA.  
The gap at $z\sim 3.84$ is due to interstellar \ion{Na}{1}; the pixels 
affected by this line (at 5894.6~\AA) were also removed from our analysis.  
Second, there is a more gradual and extended dip in counts around 
$z\sim 2.8-3.4$.  The Ly$\alpha$ emission line passes from the $g^*$ to 
the $r^*$ band at $z\sim 3.5$.  Therefore, the observed colors of QSOs 
change relatively rapidly in this regime, and so one might worry that 
the color--based algorithm which SDSS uses for targetting QSO candidates 
for observation is less accurate at these redshifts.  In particular, one 
might worry that the dip in counts evident in Figure~\ref{fig:hist} 
signals the fact that the selection algorithm chooses a biased subset 
of the complete population at these redshifts.  This is of particular 
concern because the feature in $\tau_{\rm eff}(z)$ occurs in this redshift
range.

A detailed discussion of the effect of the color--based selection is 
presented in Appendix~\ref{seffects}, which argues that the selection 
does not result in a biased measurement of the mean transmission.  
It also argues that, if there are inaccuracies in how the SDSS 
spectrograph is calibrated, they do not give rise to a feature in the 
evolution of the mean transmission.  

\subsection{The ratio of signal-to-noise}\label{snratio}

\begin{figure}[t]
\plotone{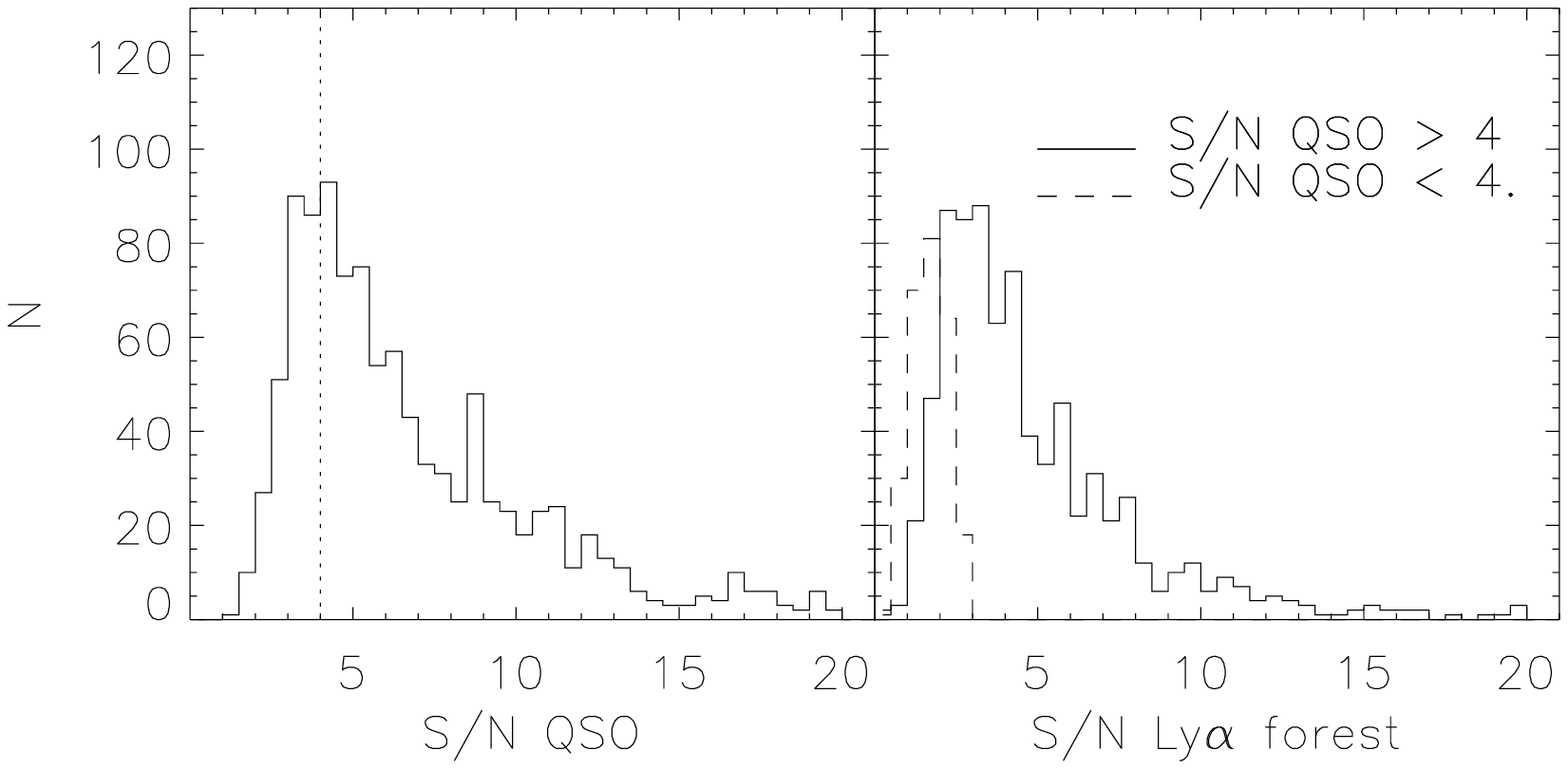}
\vspace{-9cm}
\caption{Distribution of signal-to-noise ratios in our sample 
redward (left) and blueward (right) of the Ly$\alpha$ emission 
line.  Solid and dashed histograms in the panel on the right are for 
spectra with $S/N$ ratios (computed redward of the Ly$\alpha$ emission
line) which are greater than and less than 4. }
\label{s/n}
\end{figure}

\begin{figure}
\plotone{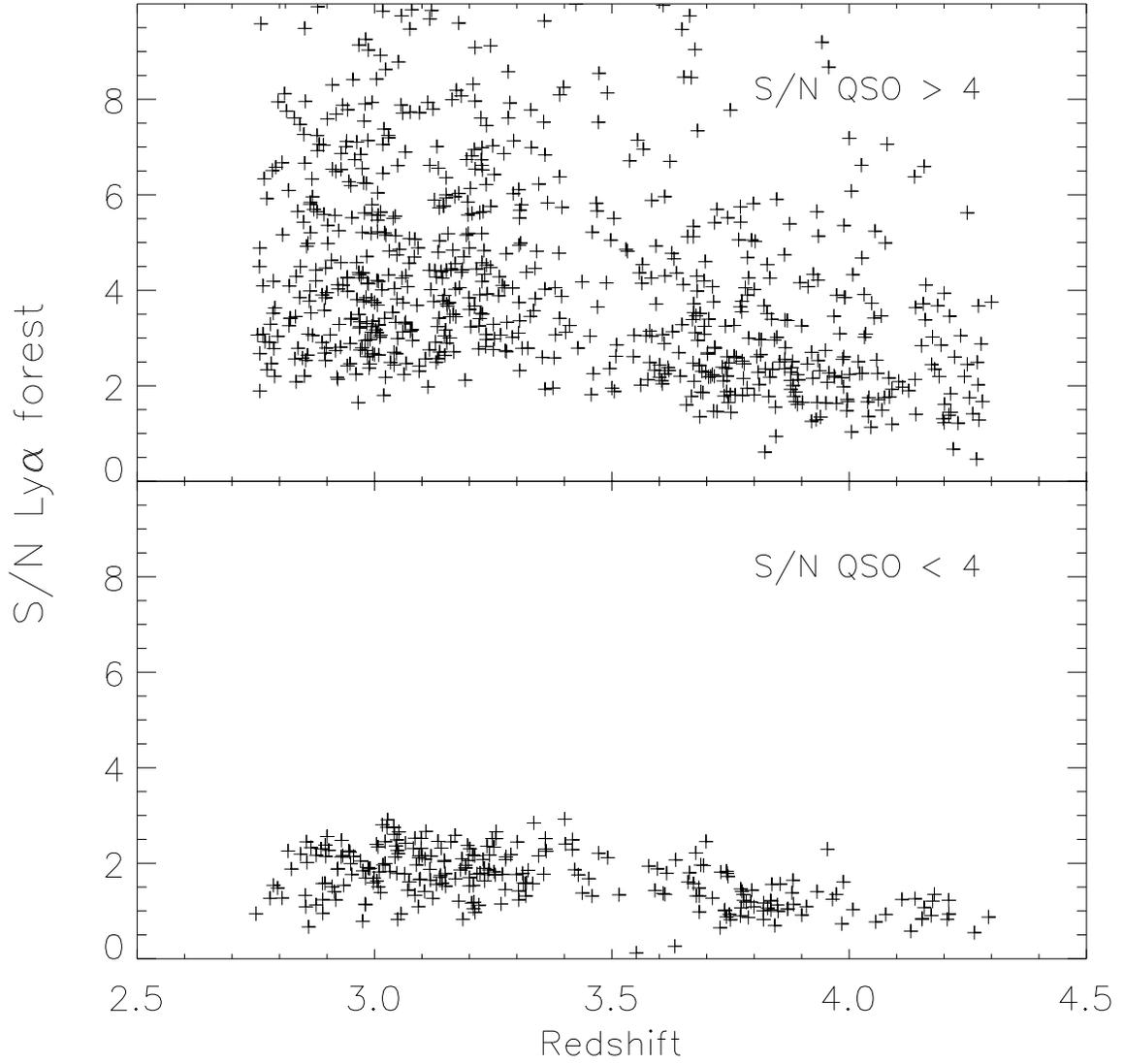}
\caption{Signal-to-noise ratios in the Ly$\alpha$ forest as a function 
of redshift.  Top and bottom panels show results for spectra with 
larger/poorer ratios redward of $\lambda_\alpha$.   }
\label{lyasnz}
\end{figure}

\begin{figure}
\plotone{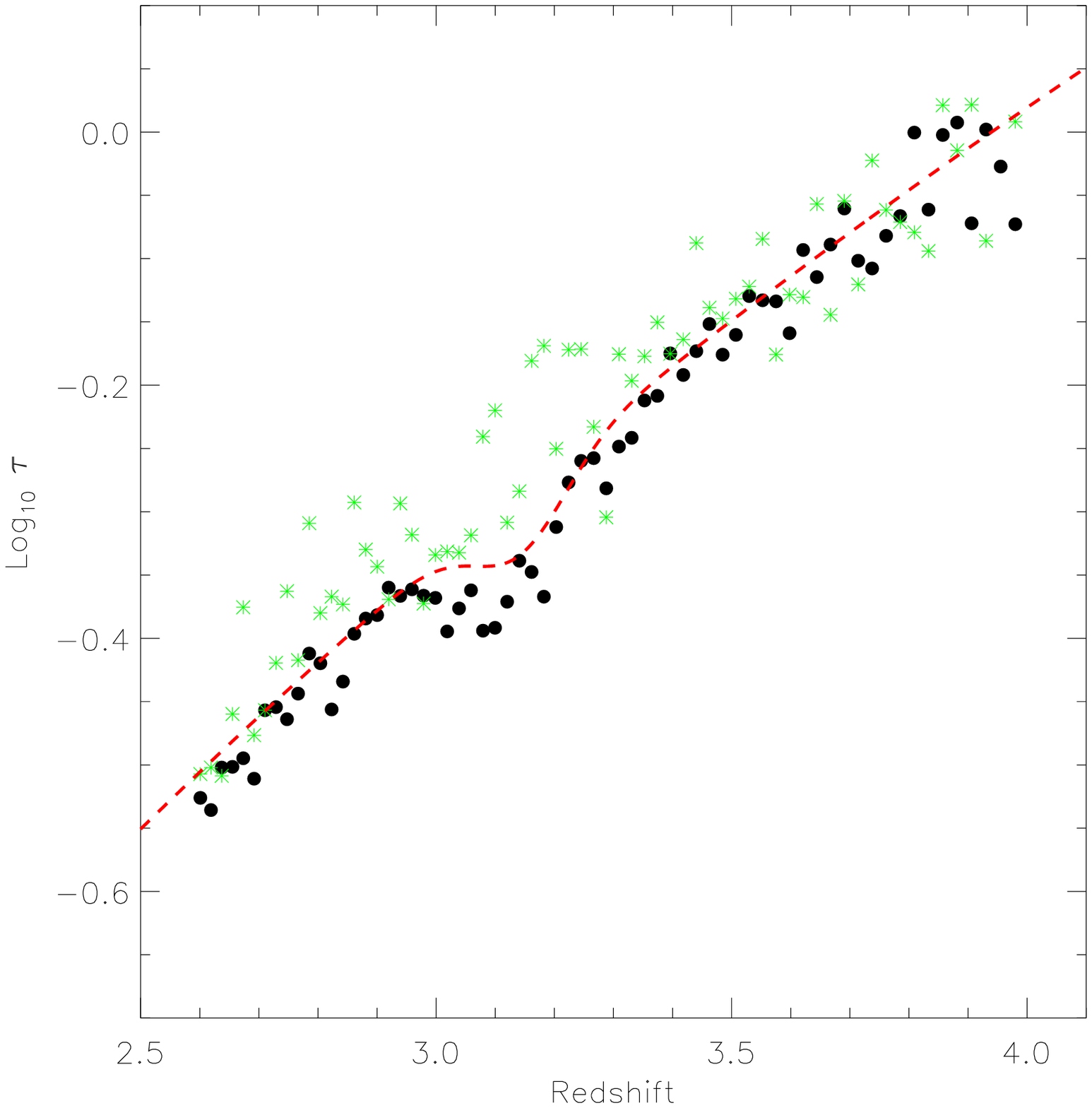}
\caption{Evolution of the effective optical depth estimated using 
spectra which have low (stars) and high (circles) signal-to-noise ratios 
redward of the Ly$\alpha$ emission line.  Dashed line (same as in 
Figure~\ref{meantau}) shows the evolution inferred for the entire sample.  
The feature in $\tau_{\rm eff}$ is more obvious in the spectra 
which have larger signal-to-noise ratios.}
\label{hilosn}
\end{figure}

The signal-to-noise ratio in our sample is low.  We would like to be 
sure that the evolution of $\tau_{\rm eff}$ does not depend on $S/N$.  
The panel on the left of Figure~\ref{s/n} shows the distribution of the 
typical $S/N$ when the ratio is computed on the red side of $\lambda_\alpha$ 
in each spectrum.  The panel on the right shows the distribution of typical 
$S/N$ ratios in the Ly$\alpha$ forest region of each spectrum for QSOs with 
large (solid) and small (dashed) $S/N$ ratios redward of $\lambda_\alpha$.  

Figure~\ref{lyasnz} shows the distribution of $S/N$ ratios in the forest 
as a function of redshift.  The two panels are for spectra with $S/N>4$ 
and $S/N<4$ redward of the Ly$\alpha$ emission line.  The upper panel 
shows that the higher redshift spectra tend to have lower $S/N$ ratios.  
Comparison with the typical noise curves in different panels of 
Figure~\ref{fig:composite} suggests that the noise in the Ly$\alpha$ 
forest is approximately the same at all redshifts.  If the noise does 
not change with redshift, then the fact that the mean transmission is 
smaller at high redshift means that the typical $S/N$ ratios will also 
be smaller at high redshift.  This is qualitatively consistent with 
the trend in Figure~\ref{lyasnz}.  
Therefore, Figure~\ref{lyasnz} suggests that if we keep only spectra with 
larger values of $S/N$ (i.e., we use only those spectra which contribute 
to the top panel), then we will not introduce any severe redshift dependent 
cuts into the sample.  An estimate of $\tau_{\rm eff}(z)$ in the higher 
signal-to-noise sample should therefore be fair.  Note that it is important 
to make this cut using the $S/N$ ratio redward of Ly$\alpha$; if the noise 
is approximately the same for all spectra, then eliminating spectra with 
small $S/N$ ratios in the Ly$\alpha$ forest region biases the sample 
towards larger transmission.  

Figure~\ref{hilosn} shows the evolution of the effective optical depth 
estimated using spectra which have low (stars) and high (circles) 
signal-to-noise ratios redward of the Ly$\alpha$ emission line.  
Dashed line (same as in Figure~\ref{meantau}) shows the evolution inferred 
for the entire sample.  There are many fewer low $S/N$ spectra, so the 
stars scatter wildly.  In contrast, the feature in $\tau_{\rm eff}$ is 
more obvious in the spectra which have S/N$>4$ (796 of the 1061 spectra 
in our full sample form this higher S/N subsample).  
The plots which follow show results from the higher $S/N$ subsample only.  

\subsection{Dependence on smoothing scale}
The $\chi^2$ estimate of $\tau_{\rm eff}$ comes from a sum over the fluxes 
in each pixel, so it can be thought of as an estimate which smoothes the 
data as little as possible.  It is interesting to see if the inferred 
evolution depends on how the measurement is smoothed.  For example, we 
could have chosen to compute the mean transmission averaged over the 
spectrum of each QSO:  i.e., we could average the transmission over all 
the Ly$\alpha$ forest pixels in the spectrum of each object, and plot it 
as a function of the mean redshift of the forest (recall that this 
redshift depends on the redshift of the QSO).  Or we could split the 
Ly$\alpha$ forest of each spectrum into two pieces, or four pieces, or 
eight, etc., down to the minimum possible scale which is set by the SDSS 
pixel size, and plot the mean transmission as a function of the mean 
redshift in the half-spectrum, the quarter-spectrum, etc.  
There is no compelling reason for prefering one choice to another since, 
whatever sets the physical scale in the forest (e.g., the Jeans smoothing 
scale is expected to be about 30 km s$^{-1}$), the SDSS spectrograph does 
not resolve it.  

\begin{figure}
\plotone{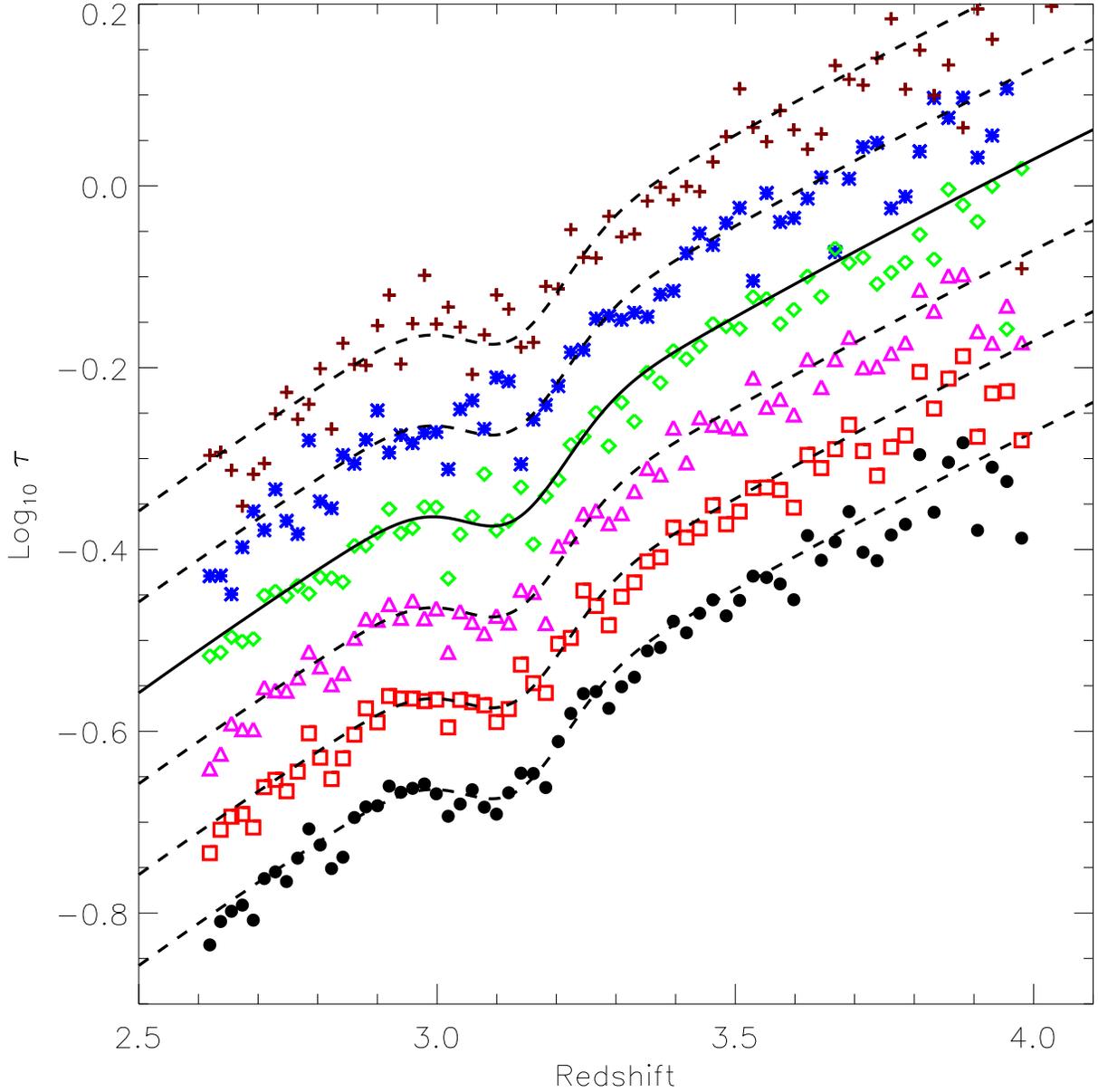}
\caption{Evolution of the effective optical depth in the Ly$\alpha$ 
forest:  dependence on smoothing scale.  Solid curve shows the evolution 
derived from applying the $\chi^2$ technique on the $S/N>4$ sample; 
dashed curves show the same, except that they have been offset from 
the solid curve for clarity.  Symbols, which have also been offset for 
clarity, show the mean $\tau_{\rm eff}$ derived from cutting the spectra 
in half (top), in quarters (second from top), in eight (third from top), 
and so on, down to 64 pieces, and plotting versus the median redshift 
$z_\alpha$ in the half-spectrum, the quarter-spectrum, etc.  Evidence for 
a feature in $\tau_{\rm eff}$ becomes apparent only when the size over 
which the measurement is averaged is smaller than the size of the feature.}
\label{fig:binsize}
\end{figure}

\begin{figure}
\plotone{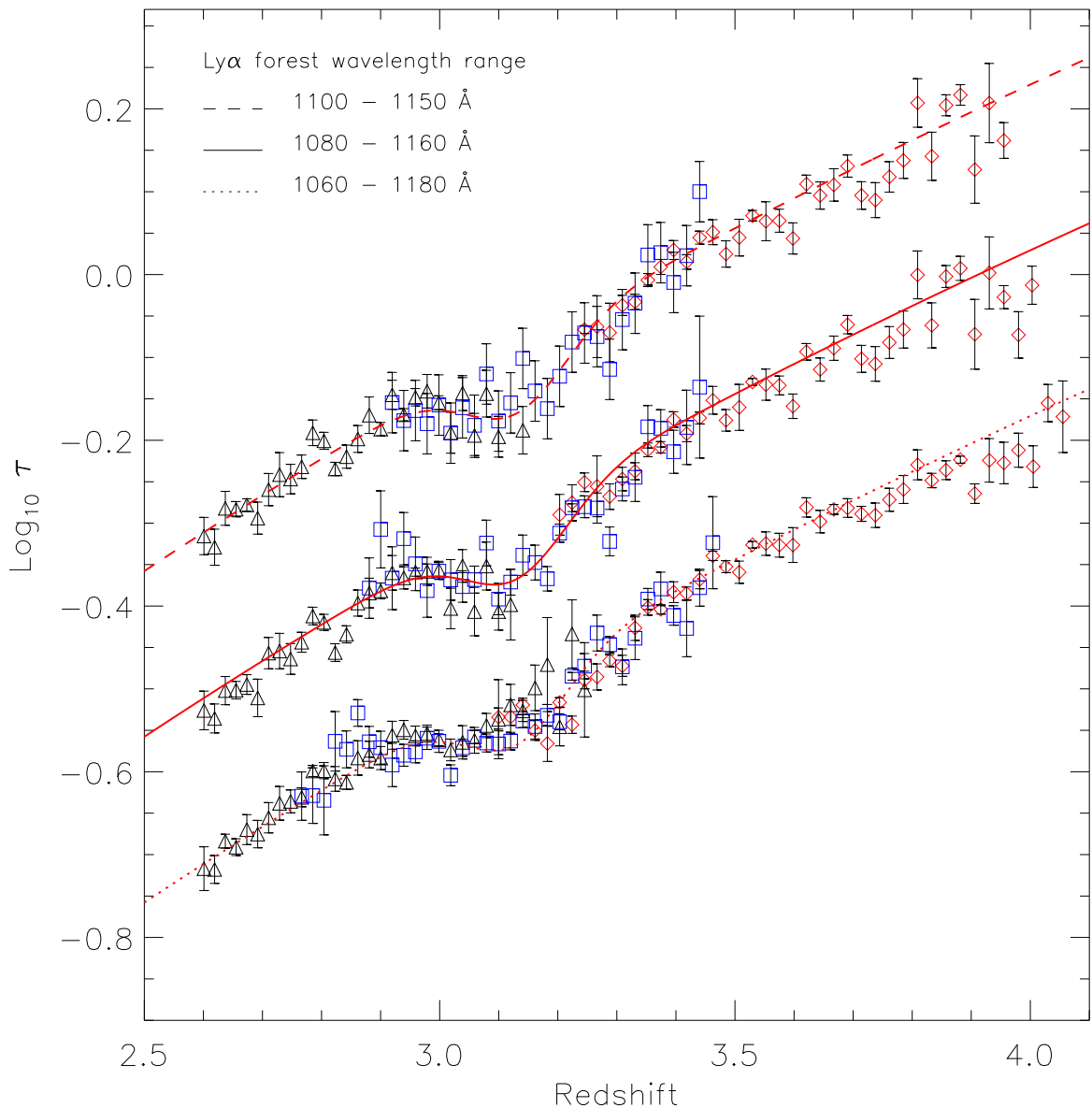}
\caption{Dependence of the effective optical depth on the definition of 
the Ly$\alpha$ forest. 
Different sets of curves and symbols show results for three definitions 
of the wavelength range spanned by the Ly$\alpha$ forest:
$1060-1180$~\AA\ (bottom), $1080-1160$~\AA\ (middle), and 
$1100-1150$~\AA\ (top). The upper and lower sets have been shifted 
upwards and downwards by 0.2 in $\log_{10}\tau$. 
Smooth line shows the evolution of $\tau_{\rm eff}$ determined by the 
$\chi^2$ technique applied to the $S/N>4$ sample (see Table~1); 
dashed and dotted curves show the same, except that they have been offset
upwards and downwards by 0.2 in $\log_{10}\tau$. 
Triangles, squares and diamonds show measurements from QSOs at 
low ($z<3.2$), medium ($3.2<z<3.7$) and high ($z>3.7$) redshifts.  
Error bars were computed by bootstrap re-sampling as described in 
the text. }
\label{fig:MBtau4}
\end{figure}

\begin{figure}
\plotone{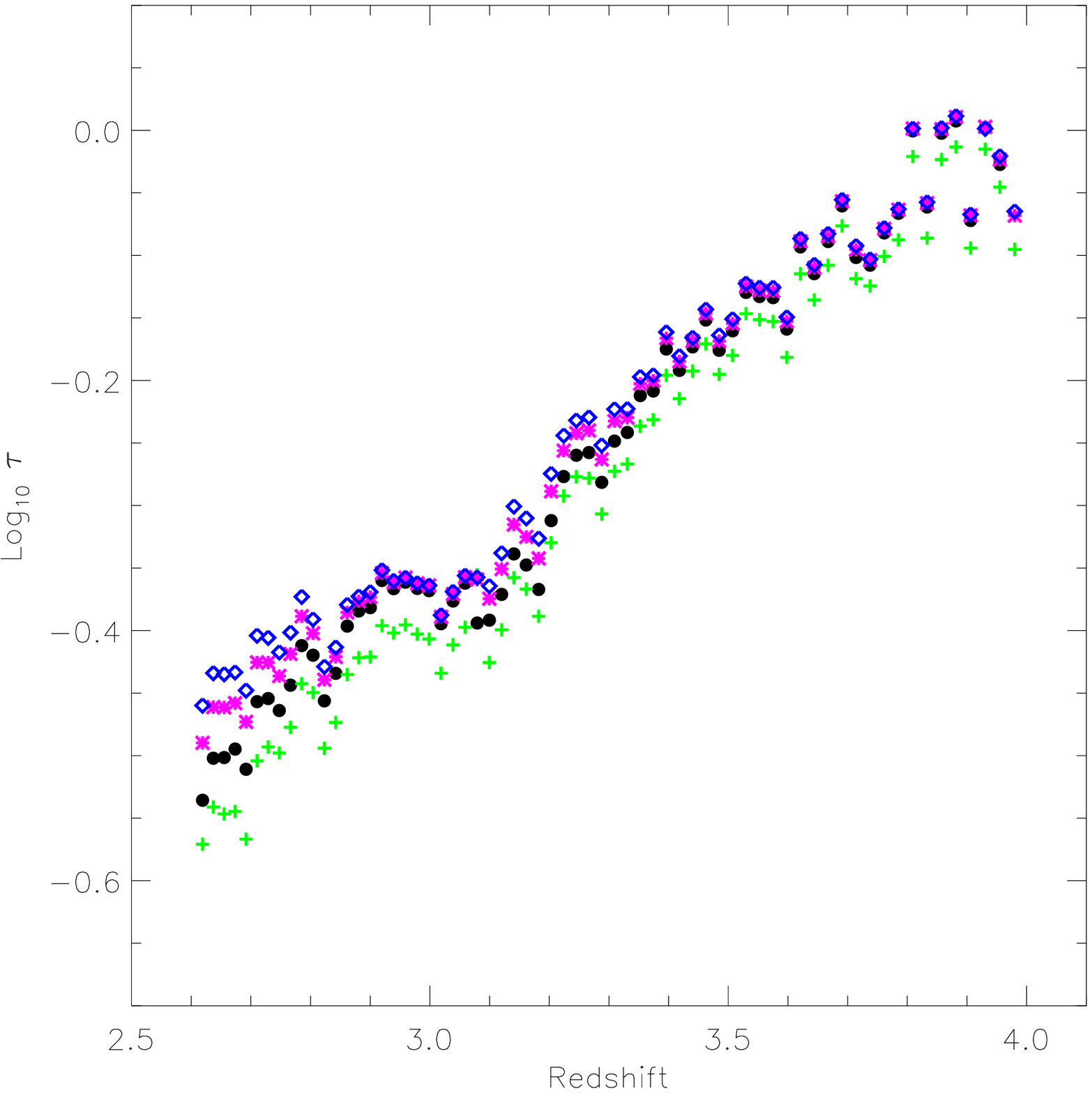}
\caption{Evolution of the effective optical depth in the Ly$\alpha$ forest: 
dependence on continuum shape and normalization.  Filled circles show 
our standard, crosses result from assuming a featureless power-law, 
diamonds from retaining features, but normalizing by the region in front 
of \ion{Si}{4} instead of \ion{C}{4}, and stars use both regions to 
normalize the spectra.  The feature in $\tau_{\rm eff}$ is present in all 
cases.  }
\label{taunorm}
\end{figure}

Figure~\ref{fig:binsize} shows the evolution of the effective optical 
depth in the Ly$\alpha$ forest as a function of smoothing scale. The solid 
curve shows the evolution derived from applying the $\chi^2$ technique
to the sample with higher $S/N$ (parameters are in Table~1); 
dashed curves show the same, but have been offset from 
the solid curve for clarity.  Symbols, which have also been offset for 
clarity, show the mean $\tau_{\rm eff}$ derived after cutting the spectra 
in half (top), in quarters (second from top), in eight (third from top), 
and so on, and plotting versus the median redshift $z_\alpha$ in the 
half-spectrum, the quarter-spectrum, etc.  
The figure shows that evidence for a feature 
in $\tau_{\rm eff}$ becomes apparent only when the size over which the 
measurement is averaged is smaller than the size of the feature.  
Once the smoothing scale is smaller than 
$\Delta\lambda_{obs}\approx 40\AA$, i.e., about 3,000 km~s$^{-1}$, 
the feature in $\tau_{\rm eff}$ is robust.  

Although the evolution of $\tau_{\rm eff}$ inferred from mean value 
statistics does not depend on the bin size, the estimated evolution from 
median value statistics does.  This is a signature that the underlying 
distribution of flux decrements is non-Gaussian.  [A non-Gaussian 
distribution is not unexpected; it is seen in hydro-dynamical simulations 
of the Ly$\alpha$ forest, and there are theoretical models relating it to 
the non-Gaussian distribution of mildly nonlinear density fluctuations 
(e.g., Gazta\~naga \& Croft 1999).]  
Appendix~\ref{app:skew} summarizes the effects of using median rather 
than mean value statistics to make all our estimates.  It shows that, 
for the median as for the mean, a feature in $\tau_{\rm eff}(z)$ 
appears at $z\sim 3.2$.

\subsection{Dependence on definition of forest}

Figure~\ref{fig:MBtau4} shows that the feature in $\tau_{\rm eff}$ is not 
caused by QSOs in one particular redshift range, nor does it depend on 
the precise wavelength range used to define the forest.  
The different sets of curves show results for three different choices of 
the wavelength range spanned by the Ly$\alpha$ forest:  
the middle curve and associated symbols show results for the wavelength 
range $1080-1160$~\AA; the upper and lower curves, which have been shifted 
by $\log_{10} \tau=\pm 0.2$ for clarity, show results for the wavelength 
ranges $1100-1150$~\AA\ and $1060-1180$~\AA, respectively.  Results for 
the larger range are more likely to be affected by inaccuracies in our 
continuum fit which arise from the fact that the Ly$\alpha$ emission line 
at $\lambda_\alpha=1215.67$~\AA\ has a tail which extends to shorter 
wavelengths (cf. Figures~\ref{compz} and~\ref{prox}).  
The shortest wavelength range is more conservative about the accuracy 
of the continuum fit in the vicinity of the emission line.  

In each set of curves, triangles, squares and diamonds show 
$\log_{10} \tau_{\rm eff}$ estimated from the mean transmission in the 
pixels of spectra of QSOs in the redshift ranges $z<3.2$, $3.2<z<3.7$, 
and $z>3.7$.  The figure shows that the measurements from the three 
redshift ranges fit smoothly onto each other and overlap (the amount of 
overlap depends, of course, on the wavelength range the Ly$\alpha$ forest 
spans), even though the SDSS sample in the middle redshift range is 
incomplete (Appendix~\ref{seffects}).  
Also, note that the feature in $\tau_{\rm eff}$ does not depend 
on the wavelength range used to define the Ly$\alpha$ forest---although 
the dip is perhaps more obvious in our most conservative definition of 
the forest (top curve) than when the forest overlaps the tails of 
Ly$\alpha$ emission line (bottom).  

\subsection{Dependence on the shape and normalization of the continuum}

One of the novel features of the continuum we use is the incorporation 
of non-power-law features (such as the bumps at $\sim 1070$~\AA\ and 
$\sim 1120$~\AA).  Other studies of $\tau_{\rm eff}$ using similarly 
low resolution spectra (e.g. Schneider, Schmidt \& Gunn et al. 1991) do 
not incorporate these features.  To compare our results with theirs, we 
repeated the analysis assuming that the continuum was a featureless 
power-law with slope $\alpha_\lambda=-1.56$ (see, e.g., the smooth 
solid lines in Figure~\ref{fig:spectra}).  

Filled circles in Figure~\ref{taunorm} show $\tau_{\rm eff}(z)$ for 
our standard definition of the continuum, and crosses show the result 
of using the featureless power-law.  Since the power-law continuum has less 
flux than our standard, the inferred mean transmission is always higher, so 
the associated $\tau_{\rm eff}$ always slightly lower.  Retaining features 
in the continuum, but normalizing by the flux in the range $1350-1370$~\AA\ 
(the region just blueward of the \ion{Si}{4} emission line) instead of 
$1450-1470$~\AA\ (diamonds in Figure~\ref{taunorm}), or normalizing by the 
flux in both regions (stars in Figure~\ref{taunorm}), makes little 
difference at high redshifts, but begins to matter at lower redshifts.  
A glance at the composite spectra in Figure~\ref{fig:composite} shows 
why:  a power-law which is normalized to fit the range 
$1450-1470$~\AA\ only provides a good fit to the region in front of 
\ion{Si}{4} at higher redshifts, but systematically underestimates the 
flux there by a small amount at lower redshifts.  Changing the normalization 
increases the flux in the continuum, which reduces the inferred transmission 
and increases the effective optical depth.  
Although the evolution of $\tau_{\rm eff}$, particularly at lower redshifts 
does depend on how we normalize the spectra, the feature in $\tau_{\rm eff}$ 
is present, at the same redshift, in all cases.  

We have also examined (and excluded) the possibility that the feature is
produced by intrinsic absorption in a subset of QSO in our sample by 1)
determining that the optical depth is not a function of velocity of the
absorber from the emission redshift, and 2) determining that the
effect is not influenced by the well-known velocity shift of \ion{C}{4} 
emission (e.g. Richards et al. 2002b).

\section{Discussion}
When applied to a sample of 1061 QSO spectra drawn from the SDSS database, 
two methods---which solve simultaneously for the shape of the QSO continuum 
and for the evolution of the mean transmission---give consistent results.  
Both methods show that the continuum in the wavelength range between 
the Ly$\alpha$ and Ly$\beta$ emission lines is not smooth, but has 
features in it.  The two methods also show that although the effective 
Ly$\alpha$ optical depth decreases smoothly with time, 
$\tau_{\rm eff}\propto (1+z)^{3.8\pm 0.2}$, it drops by about 
10 percent from $z\sim 3.3$ to $z\sim 3.1$, and it recovers to the 
original smooth scaling by $z\sim 2.9$.  

\begin{figure}
\plotone{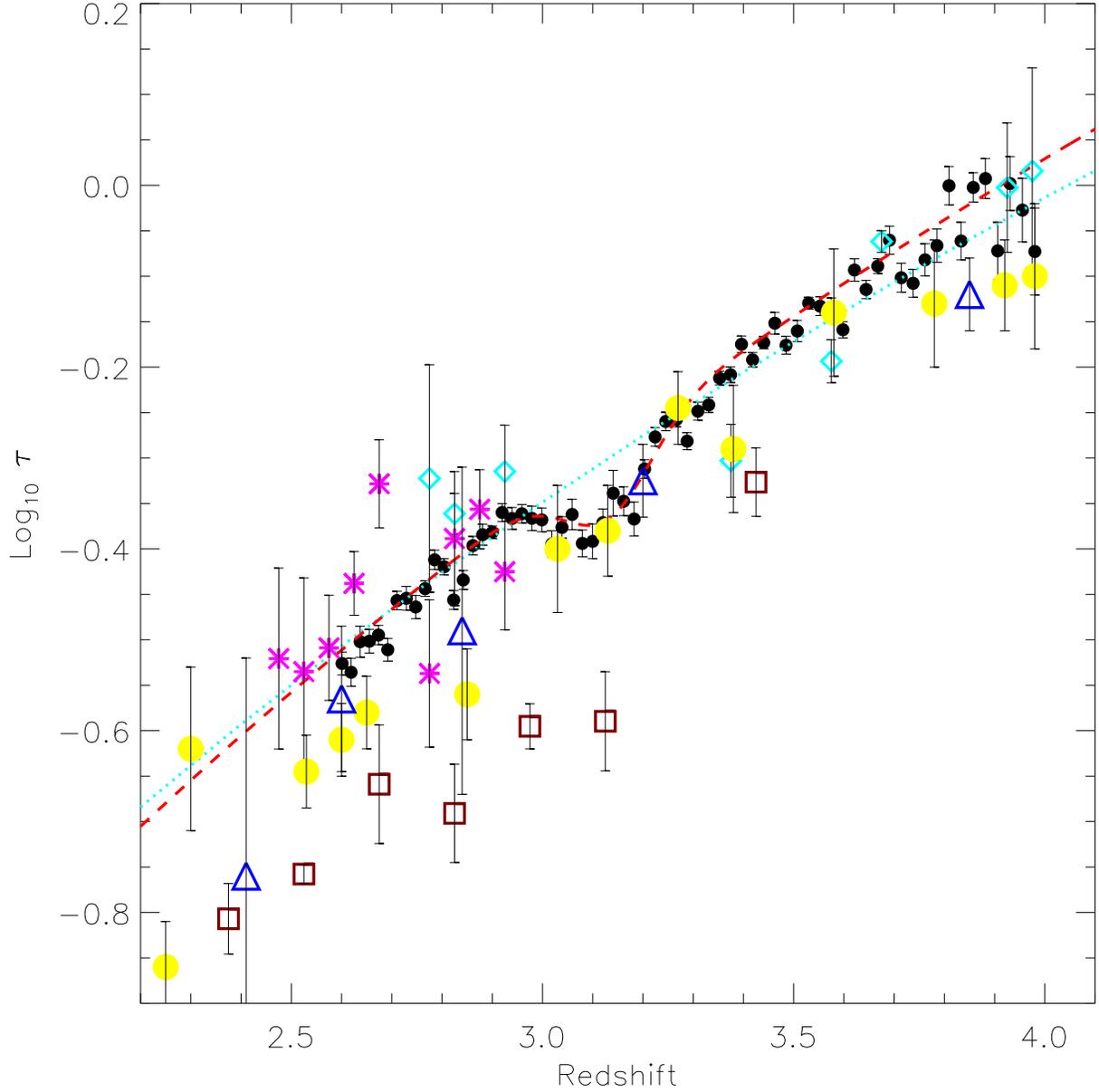}
\caption{Comparison of measurements of the evolution of the effective 
optical depth in the Ly$\alpha$ forest.  Stars, diamonds, squares and 
small filled circles show measurements from 42 low resolution spectra by 
Sargent, Steidel \& Bocksenberg (1989), 
33 from Schneider, Schmidt \& Gunn (1991), 
42 from Zuo \& Lu (1993), and the subset of 796 QSOs in the SDSS sample 
which had $S/N>4$ and were studied in this paper.  
Triangles and large filled circles show measurements in $\sim 10$ higher 
resolution spectra by McDonald et al. (2000) and Schaye et al. (2000).  
Dotted line shows the evolution reported by Press, Rybicki \& 
Schneider (1993), and dashed line shows the evolution given in Table~1.}
\label{taulit}
\end{figure}

A comparison of our measurement of $\tau_{\rm eff}$ with the findings of 
other authors is shown in Figure~\ref{taulit}.  Stars, diamonds, squares 
and small filled circles show measurements from 
low resolution spectra of 42 QSOs by Sargent, Steidel \& Bocksenberg (1989), 
33 QSOs from Schneider, Schmidt \& Gunn (1991), 
42 QSOs from Zuo \& Lu (1993), and 796 QSOs from the SDSS sample studied 
in this paper (the 796 spectra with $S/N>4$ out of the full sample of 
1061 QSOs; the Ly$\alpha$ forest was defined to span the 
range $1080-1160$~\AA).
Dotted line shows the evolution in the Schneider, Schmidt \& Gunn sample 
reported by Press, Rybicki \& Schneider (1993), and dashed line shows the 
evolution given in Table~1.  Large filled circles and open triangles 
show measurements from high resolution high signal-to-noise spectra by 
Schaye et al. (2000) and McDonald et al. (2000)---but note that although 
the two sets of analyses differed (McDonald et al. quote results in coarser 
redshift bins than do Schaye et al.), they were performed on essentially 
the same set of $\sim 10$ QSO spectra.  

Figure~\ref{taulit} shows that our measurements are in general agreement 
with previous work based on low resolution, low signal-to-noise spectra 
(compare small filled circles with stars and diamonds).  
However, because our sample is so much larger than any previously 
available, it is possible to measure the evolution of $\tau_{\rm eff}$ in 
smaller redshift bins than was possible previously.  The figure also shows 
that our measurement of $\tau_{\rm eff}$ results in about ten percent less 
transmitted flux than suggested by recent measurements from higher 
resolution spectra (triangles and large circles).   At higher redshifts, 
where the absorption is large, it becomes increasingly difficult to 
estimate the continuum reliably for high resolution spectra. This, 
together with small number statistics may account for some 
of the discrepancy at $z>3.5$.  At lower redshifts, 
some of the discrepancy may arise because damped Ly$\alpha$ systems 
and/or metal lines, which become increasingly abundant at low redshifts, 
have been removed from the higher resolution spectra, but are still present 
and contributing to the effective optical depth in our sample.  Although 
the mean transmission measured in noisy and 
low resolution spectra may yield a biased measure of the slope and amplitude 
of the evolution of the true effective optical depth 
(e.g., Steidel \& Sargent 1987), 
it is difficult to see why this bias should lead to a 
{\it feature} in $\tau_{\rm eff}(z)$.  
Thus, whereas the slope and amplitude of the evolution we find should be 
calibrated against simulations and other measurements, we feel we have 
strong evidence that the effective optical depth of the IGM changed 
suddenly around $z\sim 3.2$.  

The gradual evolution of the effective optical depth, and the strength 
of the feature superposed on it, both have implications for how the 
temperature and the photo-ionization rate evolve (equation~\ref{eq:tau}).  
It is interesting that the feature we see in $\tau_{\rm eff}$ occurs at 
the same redshift range as the factor of two increase in temperature 
that Schaye et al. (2000) detected.  
Schaye et al. interpreted their measurement as evidence that \Hep\ 
reionized at $z\sim 3.5$.  Hydrodynamical simulations show that our 
measurement of the evolution of $\tau_{\rm eff}$ is consistent with 
this interpretation (Theuns et al. 2002).  
The simulations can also help us understand if the mean scaling 
$\tau_{\rm eff}\propto (1+z)^{3.8\pm 0.2}$ we see leads to reasonable 
values for the amplitude and evolution of $\Gamma$.  
It would be a significant accomplishment if the simulations were also 
able to reproduce the evolution of the skewed distribution around the 
mean---the latter being quantified by how the median optical depth 
depends on smoothing scale and on redshift 
(Figure~\ref{fig:MBtaumedian}).  This is the subject of ongoing work.  

The reionization of \Hep\ is expected to be proceed more gradually than 
for \ion{H}{1}.  The fact that the feature appears relatively gradually 
in our data can be used to place constraints on how patchy the onset of 
reionization was.  When the SDSS survey is complete, it will be 
possible to compile a data set which is large enough to study different 
portions of the sky separately.  This will provide an even more direct 
constraint on the homogeneity of the Universe at the epoch of \Hep\ 
reionization.

\bigskip

{\em Acknowledgments} 

This project started when Tom Theuns visited Fermilab in September 2001.  
We would like to thank him for a discussion which led to this measurement, 
and for helpful correspondence since then.   We also thank Paul Hewett, 
Patrick McDonald, Celine Peroux, and Uros Seljak.

Funding for the creation and distribution of the SDSS Archive has been 
provided by the Alfred P. Sloan Foundation, the Participating Institutions, 
the National Aeronautics and Space Administration, the National Science 
Foundation, the U.S. Department of Energy, the Japanese Monbukagakusho, 
and the Max Planck Society. The SDSS Web site is http://www.sdss.org/.

The SDSS is managed by the Astrophysical Research Consortium (ARC) for
the Participating Institutions. The Participating Institutions are The
University of Chicago, Fermilab, the Institute for Advanced Study, the
Japan Participation Group, The Johns Hopkins University, Los Alamos
National Laboratory, the Max-Planck-Institute for Astronomy (MPIA),
the Max-Planck-Institute for Astrophysics (MPA), New Mexico State
University, the University of Pittsburgh, Princeton University, 
the United States Naval Observatory, and the University of Washington.

\appendix

\section{An iterative procedure for estimating the mean continuum}
\label{method2}

We begin by assuming that the continuum is a power-law, 
\begin{equation}
 f_{cont}(\lambda_{rest}) \propto \lambda_{rest}^{\alpha_{\lambda}}, 
\end{equation}
normalized to have the same flux as that observed in the restframe 
wavelength range $1450-1470$~\AA: 
\begin{displaymath}
 \sum_{1450}^{1470}f_{obs} = \sum_{1450}^{1470} f_{cont}.
\end{displaymath}
This wavelength range lies in front of the \ion{C}{4} emission line, 
and is free of obvious emission and absorption lines 
(e.g., Press, Rybicki \& Schneider 1993).  

\begin{figure}
\plotone{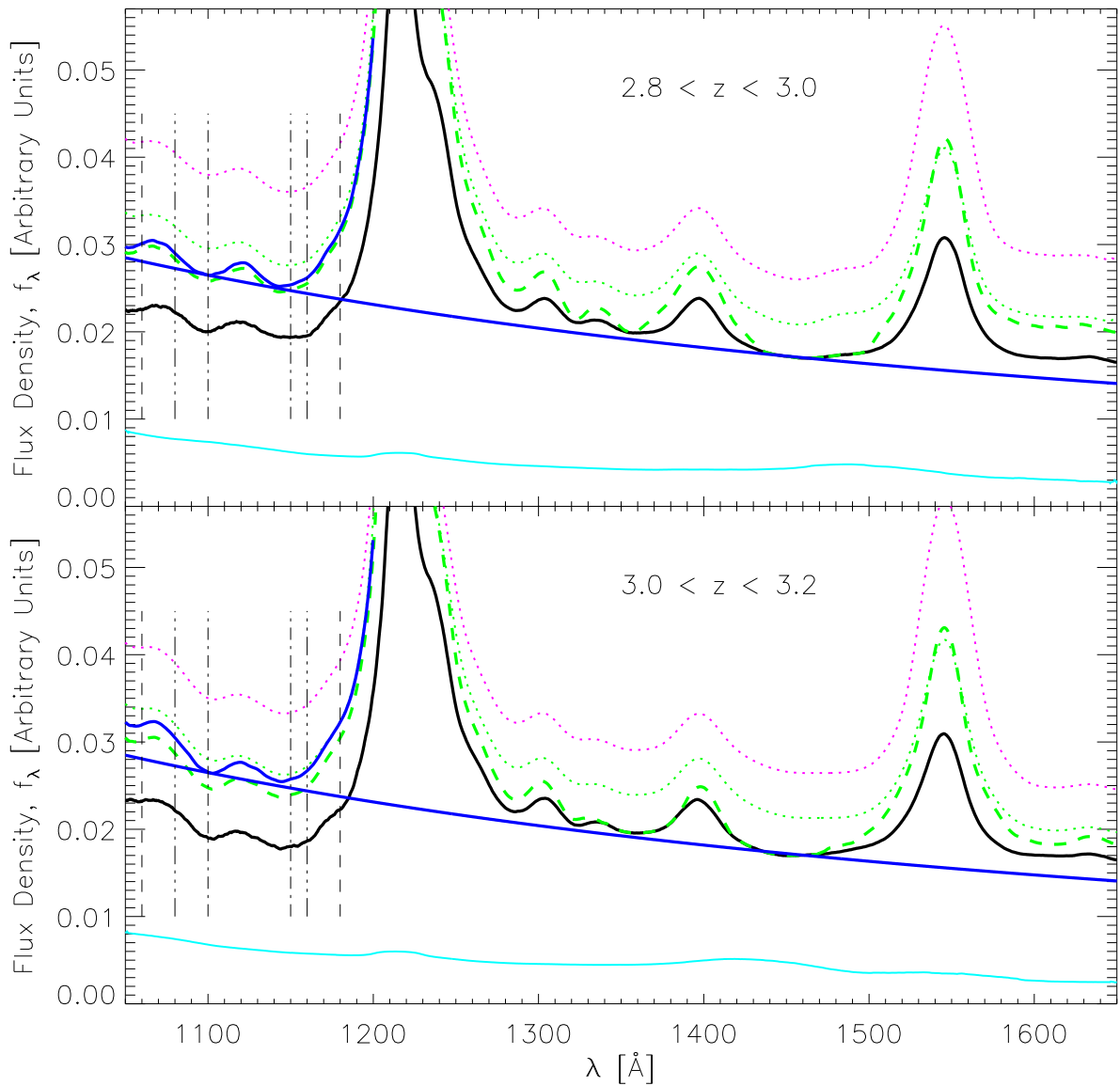}
\caption{Restframe normalized spectra (i.e. wavelengths were transformed 
to the restframe of the QSO, and the flux density of each QSO spectrum was 
normalized in the wavelength region $1450-1470$~\AA, as described in the 
text); different panels show results for different redshift bins.  
The solid curve in each panel shows the mean value of the flux 
density in all the normalized restframe spectra in the redshift bin shown.  
The lower dotted curve shows the observed rms scatter above the mean value, 
and upper dotted curve shows the 95 percentile value as a function of 
restframe wavelength.  The thin line in the bottom of each panel shows 
the typical value of the noise for individual spectra.  (The noise on each 
composite spectrum is a factor of $\sim 1/\sqrt{100}$ smaller.)  
Subtracting the noise in quadrature from the observed rms scatter (i.e., 
from the lower of the two dotted curves) gives the dashed curve.  
The curve which rises smoothly from right to left shows a power-law of 
slope $\alpha_{\lambda} = -1.56$.  
The bumpy line was obtained by simply shifting the dashed curve in the 
region shortward of $\lambda_\alpha$ upwards or downwards until its local 
minima touched the extrapolated power-law.  
The vertical lines on the left of each panel show three different 
wavelength regions adopted in defining the Ly$\alpha$ forest: 
1060--1180~\AA\ (dashed), 1080-1160~\AA\ (dot-dot-dot-dashed), 
and 1100--1150~\AA\ (dot-dashed).}
\label{fig:composite}
\end{figure}

\begin{figure}
\plotone{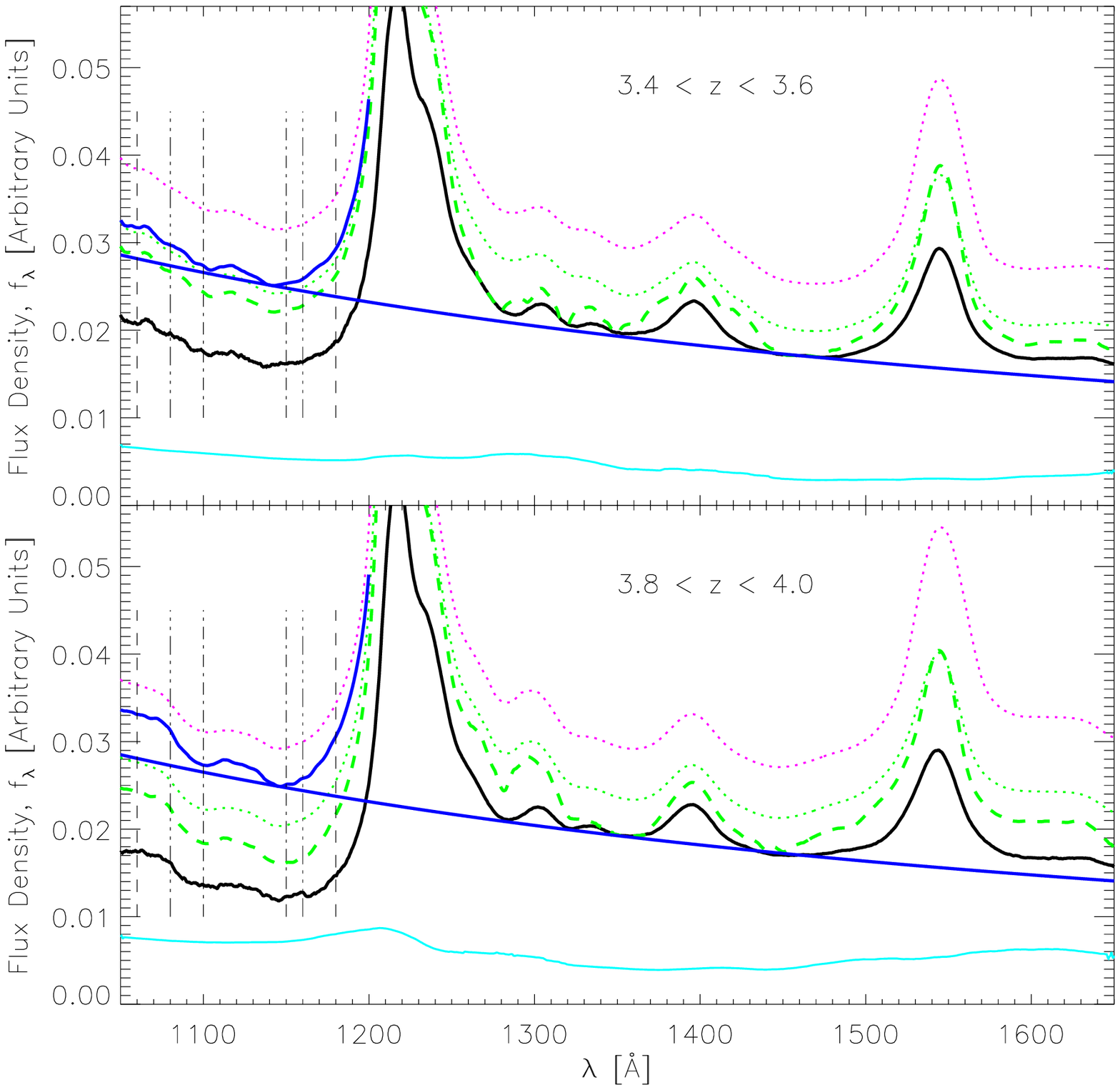}
\begin{center}
Fig. \ref{fig:composite}. -- Continued.
\end{center}
\end{figure}

From a composite spectrum of about 2,400 SDSS QSO spectra which span 
a range of redshifts, Vanden Berk et al. (2001) estimate that 
$\alpha_{\lambda} = -1.56$ in the interval $1280-5000$~\AA.  
The smooth lines in Figure~\ref{fig:spectra} show that this power-law 
continuum shape provides a reasonable description of the individual 
spectra in our sample.  Note that this slope is rather different from 
the value $\alpha_\lambda = -1.07$ used by other authors (see Vanden 
Berk et al. 2001 for further discussion).  

The different panels in Figure~\ref{fig:composite} show results obtained 
by averaging over QSOs in the redshift bins indicated in the top right 
corners.  The thick solid line in each panel shows the observed composite 
spectrum, which we will call $F_{50}$ since it is very close to the median 
value in each restframe wavelength bin.  The $\alpha_\lambda=-1.56$ 
power-law (the same in all panels) provides a reasonable fit redward of 
the Ly$\alpha$ line, but lies significantly above the composites blueward 
of $\lambda_\alpha$.  This difference is larger at higher redshift, 
qualitatively consistent with the expectation that there is more absorption 
in the forest at high redshift.  The lower dotted curve in each panel 
shows the rms scatter above the mean composite spectrum, and the upper 
dotted curve shows the curve traced out by the 95 percentile level.  
These curves provide estimates of the scatter around the mean spectrum, 
but almost all of this scatter is due to the noise.  The solid curve in 
the bottom of each panel shows the typical value of the noise:  it was 
obtained by squaring the individual noise estimates for each pixel, 
computing the average of these squared values at each bin in restframe 
wavelength, and taking the square root.  A comparison of these noise 
estimates with the observed composites shows that the typical 
signal-to-noise ratio is $\sim 5$ longward of $\lambda_\alpha=1215.67$~\AA, 
and only $\sim 3$ shortward of $\lambda_\alpha$.  The dashed lines show 
the result of subtracting the noise in quadrature before computing the 
rms scatter around the mean curve (i.e., the square root of 
 $\sum_i (f_i-F_{50})^2 - n_i^2$).  Except in the vicinity of the 
emission lines, most of the observed scatter redward of $\lambda_\alpha$ 
is due to the noise.  This is consistent with Figure~\ref{fig:pca} in 
the main text which showed that the intrinsic scatter around the 
mean continuum shape is small.  

Redward of $\lambda_\alpha$, the local minima of the dashed lines track 
the height of the mean curve, $F_{50}$, reasonably well.  If there were 
no absorption in the forest, one might expect the same to be true blueward 
of $\lambda_\alpha$.  Therefore, the next step, and the one which most 
closely parallels previous work, would be to extrapolate the power-law fit 
blueward of $\lambda_\alpha$, and use it to estimate the transmission.  
However,  as we have already seen, there appear to be emission lines in 
between the Ly$\alpha$ and Ly$\beta$ emission lines.  Using the 
$\alpha_{\lambda} = -1.56$ power-law fit blueward of $\lambda_\alpha$ and 
ignoring the bumps in the continuum will lead to biases in our estimate of 
the mean transmission.  

\begin{figure}
\plotone{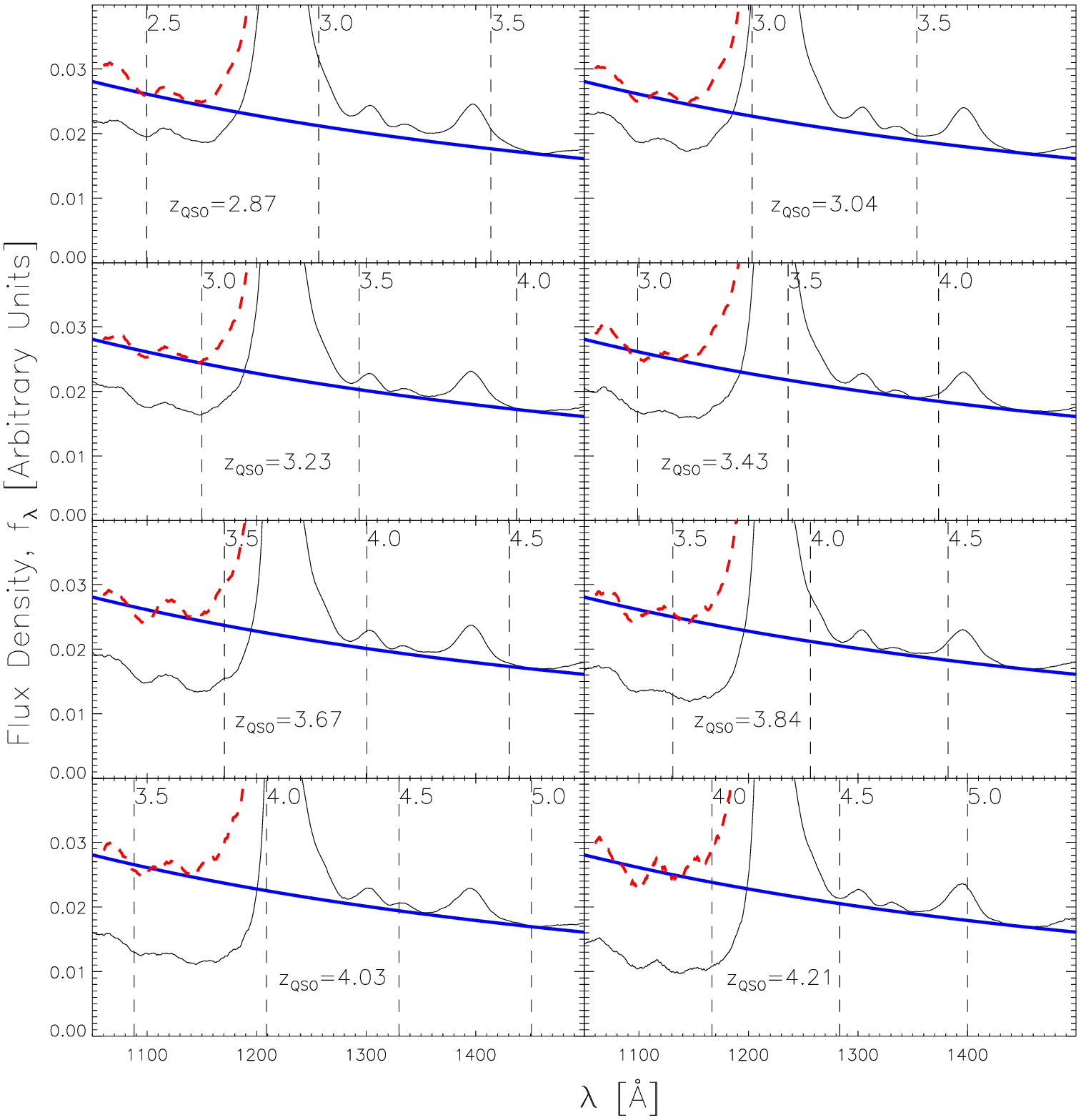}
\caption{Initial step in the iteration process.  Smooth solid line 
shows the power-law which was used as the first guess for the shape of 
the continuum.  Dashed line shows the result of using the power-law 
continuum to estimate the mean transmission, then using the mean 
transmission to correct the observed fluxes, and computing a composite 
using these corrected fluxes.  The large differences between the 
initial solid and final dashed curves indicates that the method has not 
yet converged.  The actual observed composite is the thin solid line 
at the bottom of each panel.  Labels indicate the median redshift of 
the QSOs in each panel, and the vertical lines show the asssociated 
Ly$\alpha$ forest redshifts.  }
\label{fig:iter}
\end{figure}

\begin{figure}
\plotone{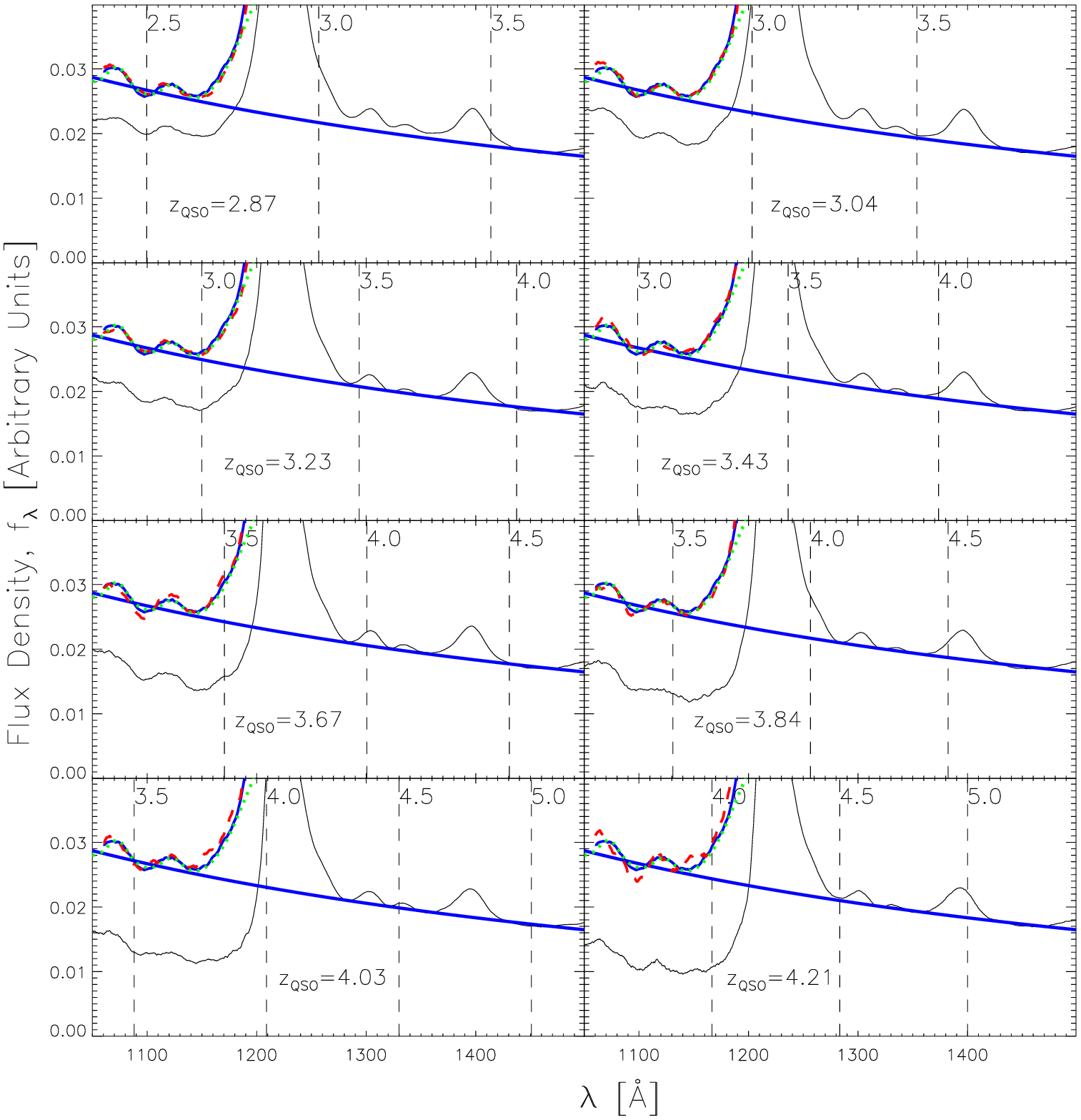}
\setcounter{figure}{14}
\caption{Continued.  Final step in the iteration process.  
Thick solid bumpy line shows the input value of the continuum, and dashed 
line shows the final value.  Dotted line shows the continuum shape 
determined by the $\chi^2$ technique described in Section~\ref{secchi2}.  
All three estimates are in reasonable agreement in all the panels, 
they are all significantly different from the initial power-law continuum 
(smooth thick solid line), and they are very different from the actual 
observed composite spectrum (thin solid line).  }
\end{figure}

We can correct for this as follows.  The extrapolated power-law may have 
approximately the correct amplitude, but it does not have bumps.  On the 
other hand, the dashed line has bumps in it, but it almost certainly does 
not have the right amplitude.  If the locations of the Ly$\alpha$ forest 
absorption features in the restframe spectrum of one QSO are uncorrelated 
with those 
in most of the others, then if we average many spectra together, the net 
effect of the forest absorption is to remove flux from the averaged 
spectrum.  If we knew how much was removed, then we could simply add this 
amount back in to the dashed line.  A simple first estimate is to shift 
the dashed line upwards until it matches the extrapolated continuum.  
This provides an estimated continuum which has the same amplitude as the 
power-law fit but has bumps in it.  (If we were willing to assume that 
the continuum does not evolve, then we could simply do this for the 
lowest redshift bin where the shift is the smallest, and use this shape 
as the continuum at all other redshifts.) The bumpy solid curves 
which sit on top of the smooth power-laws in Figure~\ref{fig:composite}
show these estimates of the continuum.  

However, these improved estimates of the true continuum are also biased 
because, in shifting curves upwards by an amount which is independent of 
$\lambda_{obs}$, we are, in effect, ignoring the evolution of the 
optical depth over the range in $z_\alpha$ spanned by the Ly$\alpha$ 
forest (recall that this range depends on the redshift bin of the QSOs; 
Figure~\ref{fig:iter} shows this dependence explicitly).  Since our goal 
is to measure small changes in the transmission, we must account more 
carefully for this evolution.  Therefore, we have adopted the following 
iterative procedure.  

Our sample of QSOs can be thought of as a collection of pixels.  
Associated with each pixel $i$ in our sample is a normalized flux 
density $f_i$, an observed frame wavelength $\lambda_{obs,i}$, and 
a restframe wavelength $\lambda_{rest,i}$.  
The transmission associated with pixel $i$ is 
 $f_i/f_{oldcont}(\lambda_{rest,i})$, where this ratio is computed in 
the restframe, and $f_{oldcont}(\lambda_{rest})$ denotes our guess for the 
shape of the continuum.  As the initial guess for $f_{oldcont}$, we can 
use the dashed curves shown in Figure~\ref{fig:composite}, or even the 
featureless power-law.  

The mean transmission $t$ at $z_\alpha$ in the Ly$\alpha$ forest is 
estimated by summing the transmission in those pixels which have the same 
observed wavelength $\lambda_{obs}=\lambda_\alpha(1+z_\alpha)$ and dividing 
by the number of such pixels.  The estimated mean transmission $t$ is then 
used to correct the observed flux density in each pixel for the absorption 
in the forest.  That is, we make a new estimate of the continuum associated 
with each pixel:   $f_{newcont,i} = f_i/t(\lambda_{obs,i})$, where we 
have written the transmission as a function of observed wavelength rather 
than of redshift in the Ly$\alpha$ forest.  
We then compute a new composite continuum by averaging $f_{newcont,i}$ over 
all pixels which have the same value of $\lambda_{rest}$, and compare it 
with the original guess.   If the initial guess for the continuum was 
accurate, then $f_{newcont}\approx f_{oldcont}$ at all $\lambda_{rest}$.  
If not, we use the new composite as a revised estimate of the continuum 
(i.e., we set $f_{oldcont}=f_{newcont}$), and iterate until convergence is 
reached (typically about three or four interations are needed).  

Figure~\ref{fig:iter} illustrates the process.  The different panels 
show different redshift bins.  Vertical dashed lines in each panel show 
how to translate from restframe wavelength to $z_\alpha$.  The thin solid 
line near the bottom of each panel shows the observed composite spectrum.  
The thick solid line shows the $\alpha_\lambda=-1.56$ power-law 
approximation to the continuum which was used to estimate the mean 
transmission.  The dashed line shows the composite which results 
from dividing the observed fluxes by the estimated mean transmission 
and averaging.  This new composite is very different from the initial 
power-law, indicating that the procedure has not converged.  

The next set of panels show the result at convergence.  
The smooth solid line shows the same power-law as before.  
The bumpy solid line shows the guess for the continuum, and the dashed 
line shows the composite one gets by correcting all observed fluxes by 
the mean transmission computed from the bumpy continuum.  Notice that 
the solid and dashed lines are in good agreement with each other.  
They are also in good agreement with the dotted line which shows the 
shape of the continuum determined by the $\chi^2-$technique described in 
the main text.  All three curves are significantly different from the 
featureless power-law.  

Having determined the mean continuum $f_{cont}$ using this iterative 
technique, it is possible to measure the mean transmission 
$\bar F\equiv \langle f_{obs}/f_{cont}\rangle$ and therefore
$\tau_{\rm eff}\equiv -\ln \bar F$ (see Figure~\ref{meantau}).

\section{Systematic effects}\label{seffects}
This appendix studies two systematic effects which might give rise to 
the feature we see in the evolution of the effective depth, and argues 
that they do not.  

\subsection{Effect of the SDSS QSO selection algorithm}

\begin{figure}[t]
\vspace{-1.cm}
\plotone{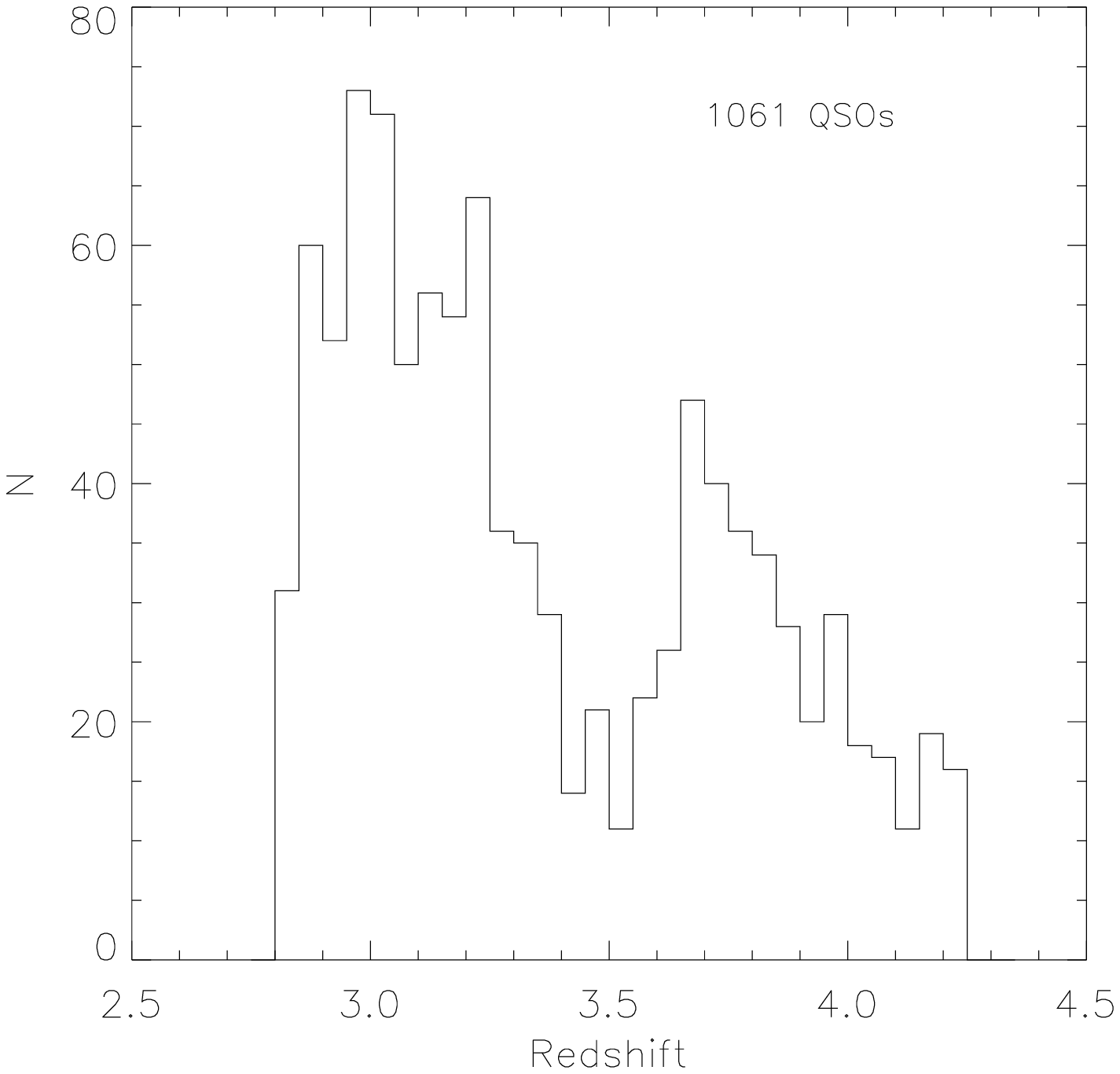}
\vspace{-3.25cm}
\caption{Distribution of QSO redshifts in the SDSS sample.  Notice 
the drop in numbers at $z\sim 3.5$.  For QSOs at these redshifts, the 
Ly$\alpha$ emission line passes from the $g^*$ to the $r^*$ filter.}
\label{fig:NzQSO}
\end{figure}

Figure~\ref{fig:NzQSO} shows the observed distribution of QSO redshifts 
in our sample.  There is an obvious drop in numbers around $z\sim 3.5$, 
which, as we describe below, is a consequence of how the colors of 
quasars change as a function of the SDSS bandpasses.  
Since the feature we see in $\tau_{\rm eff}(z)$ occurs at slightly lower 
redshifts (see Figure~\ref{meantau}), at least some of the signal comes 
from the Ly$\alpha$ forests in the spectra of these $z\sim 3.5$ QSOs.  
This Appendix studies if the feature we see in the optical depth is caused 
entirely by the QSO selection algorithm.  
Our approach is to simulate a sample of QSO spectra in which there is 
no feature in $\tau_{\rm eff}(z)$, select the subset of objects which 
SDSS would have identified as QSOs, and measure $\tau_{\rm eff}(z)$ in 
the subset.  We then check if the SDSS--selected subset shows any feature 
in $\tau_{\rm eff}(z)$.  

The algorithm used by the SDSS collaboration to target QSO candidate 
objects is described by Richards et al. (2002a).  In essence, it uses 
a stellar locus outlier rejection algorithm, further supplemented with 
a combination of cuts in the $u^*g^*r^*$, $g^*r^*i^*$, and $r^*i^*z^*$ 
color-spaces.  
To test the effects of this selection procedure, we must generate a set 
of mock QSOs for which redshifts, spectra and colors are known.  
To generate such a sample using the observed one (which is almost 
certainly incomplete in the range $3.2 < z < 3.7$) we must make some 
assumptions which we describe below.  

To generate redshifts of what we will call the complete sample, we draw 
a straight line in Figure~\ref{fig:NzQSO} from the observed number at 
$z=3.2$, $N_{\rm obs}(z=3.2)$, to the observed number, 
$N_{\rm obs}(z=3.7)$, at $z=3.7$.  
We then assume that the $N(z)$ distribution of a complete sample would 
follow $N_{\rm obs}(z)$ over the ranges $z<3.2$ and $z>3.7$, and would 
follow the smooth straight line we drew for the redshifts in between.  
Note that this means we are assuming that the observed sample is complete 
at redshifts $z>3.7$, and also at $z<3.2$.  We then generate a distribution 
of redshifts which follows this model for the complete $N(z)$ distribution.  

The SDSS QSO selection is based on color, so our next step is to assign 
colors to our mock QSOs which are consistent with their redshifts.  
We do this as follows.  Let $z_{\rm sim}$ denote the redshift of a mock 
QSO.  We randomly choose one of the observed QSOs with $z_{\rm obs}>3.7$; 
we will use it to generate a mock QSO spectrum.  The requirement that 
$z_{\rm obs}>3.7$ insures that none of the QSOs we use to generate our 
complete sample is from the regime in which we are most worried that the 
observed sample is incomplete.  
We then blue- or redshift the observed spectrum to the desired redshift 
$z_{\rm sim}$.  

\begin{figure}
\plotone{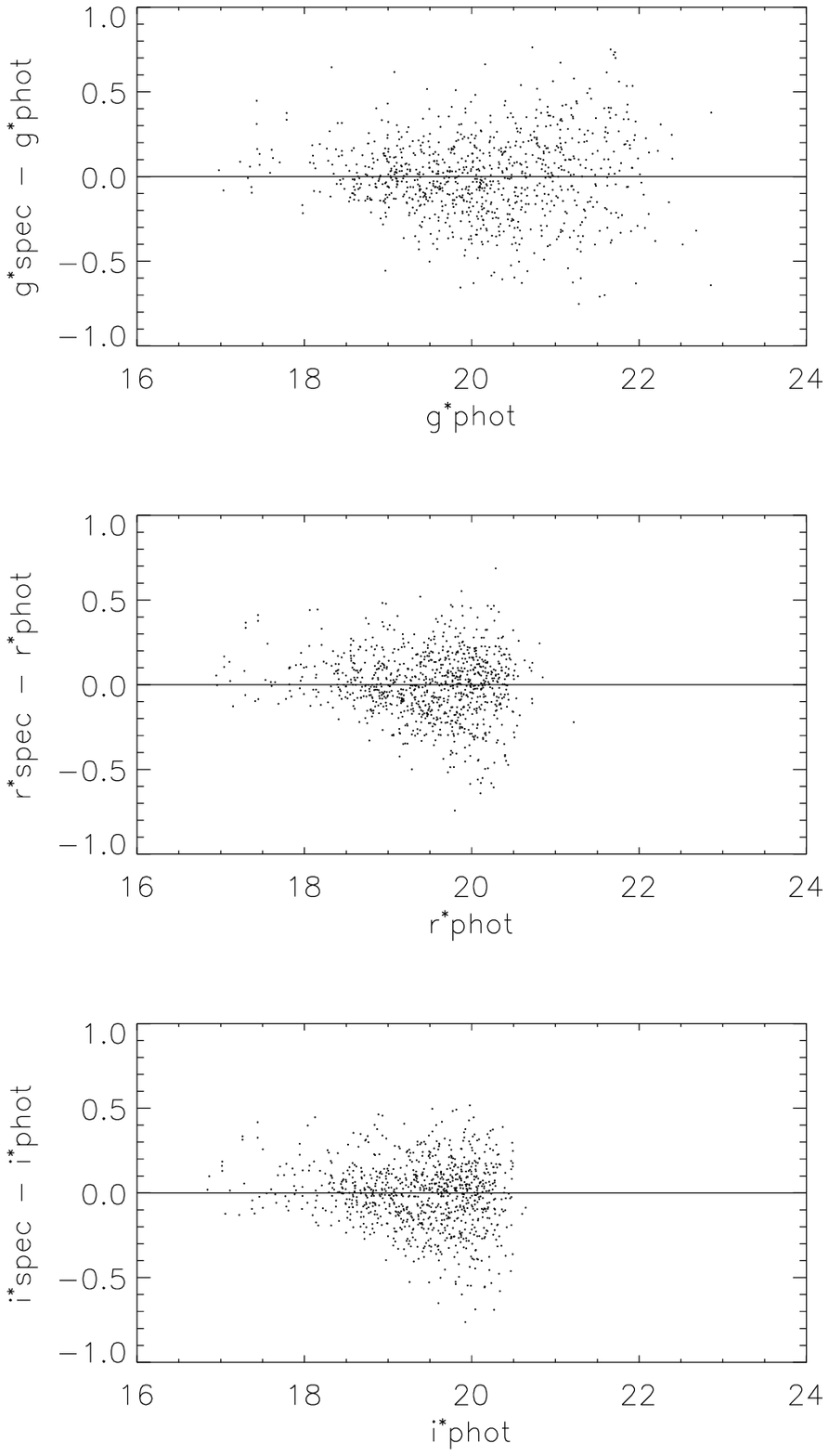}
\caption{Comparison of spec-magnitudes obtained by convolving the 
observed spectra with the SDSS filter response curves, and 
psf--magnitudes output by the photometric pipeline.}  
\label{smags}
\end{figure}

To be specific, suppose that $z_{\rm obs}>z_{\rm sim}$.  
We are interested in $D_A \equiv 1 - \langle F\rangle$, the ratio of 
the flux blueward of Ly$\alpha$ 
to that of the continuum at the same wavelength.  We know that this ratio 
evolves with redshift (see Figure~\ref{meantau}).  
We would like to generate a sample of spectra in which $\tau_{\rm eff}$ 
evolves smoothly with $z$, and so we require that $\tau_{\rm eff}$ evolve 
like a smooth line, e.g. $\tau_{\rm eff}\propto (1+z_\alpha)^{3.3}$.  
Let $D_A^{\rm fit}(z)$ denote the value of this ratio which 
gives rise to this smooth evolution in $\tau_{\rm eff}$.  
Since $\tau_{\rm eff}$ 
decreases with $z$, the higher-redshift QSO which we have blueshifted 
to $z_{\rm sim}$ has too little flux in its forest (too much absorption) 
compared to what we want.  

We remedy this as follows.  Let $\lambda_{\rm obs}$ denote the observed 
wavelength of a pixel.  It has a Ly$\alpha$ redshift 
$z_{\rm obs}^{\rm pix} = (\lambda_{\rm obs}/1215.67) - 1$.  The 
associated value of the flux decrement is $D_A(z_{\rm obs}^{\rm pix})$, 
and this may be different from $D_A^{\rm fit}(z_{\rm obs}^{\rm pix})$.  
Upon blueshifting, the Ly$\alpha$ redshift we should associate with 
the pixel is 
\begin{displaymath}
z_{\rm sim}^{\rm pix} = (\lambda_{\rm sim}/1215.67) - 1,
 \qquad {\rm where} \qquad 
\lambda_{\rm sim} = \lambda_{\rm obs}\,(1+z_{\rm sim})/(1+z_{\rm obs}).
\end{displaymath}
Therefore, after fitting the continuum to the blueshifted spectrum, 
we add 
$[D_A^{\rm fit}(z_{\rm sim}^{\rm pix})-
  D_A^{\rm fit}(z_{\rm obs}^{\rm pix})]\,f_{\rm cont}$ 
to the flux in each pixel which lies blueward of the Ly$\alpha$ emission 
line.  This ensures that the mean value of $D_A(z)$ is consistent with 
the smooth featureless evolution, but keeps approximately the same 
statistical fluctuations around the mean that were present at 
$z_{\rm obs}$.  Thus, we have a spectrum in which 
$f_{\rm sim}/f_{\rm cont}$, in the mean, follows $D_A^{\rm fit}(z)$.  
(In practice, we also add Gaussian noise with rms~$\sim 0.05$ chosen 
to be slightly smaller than that observed.)  

\begin{figure}
\plotone{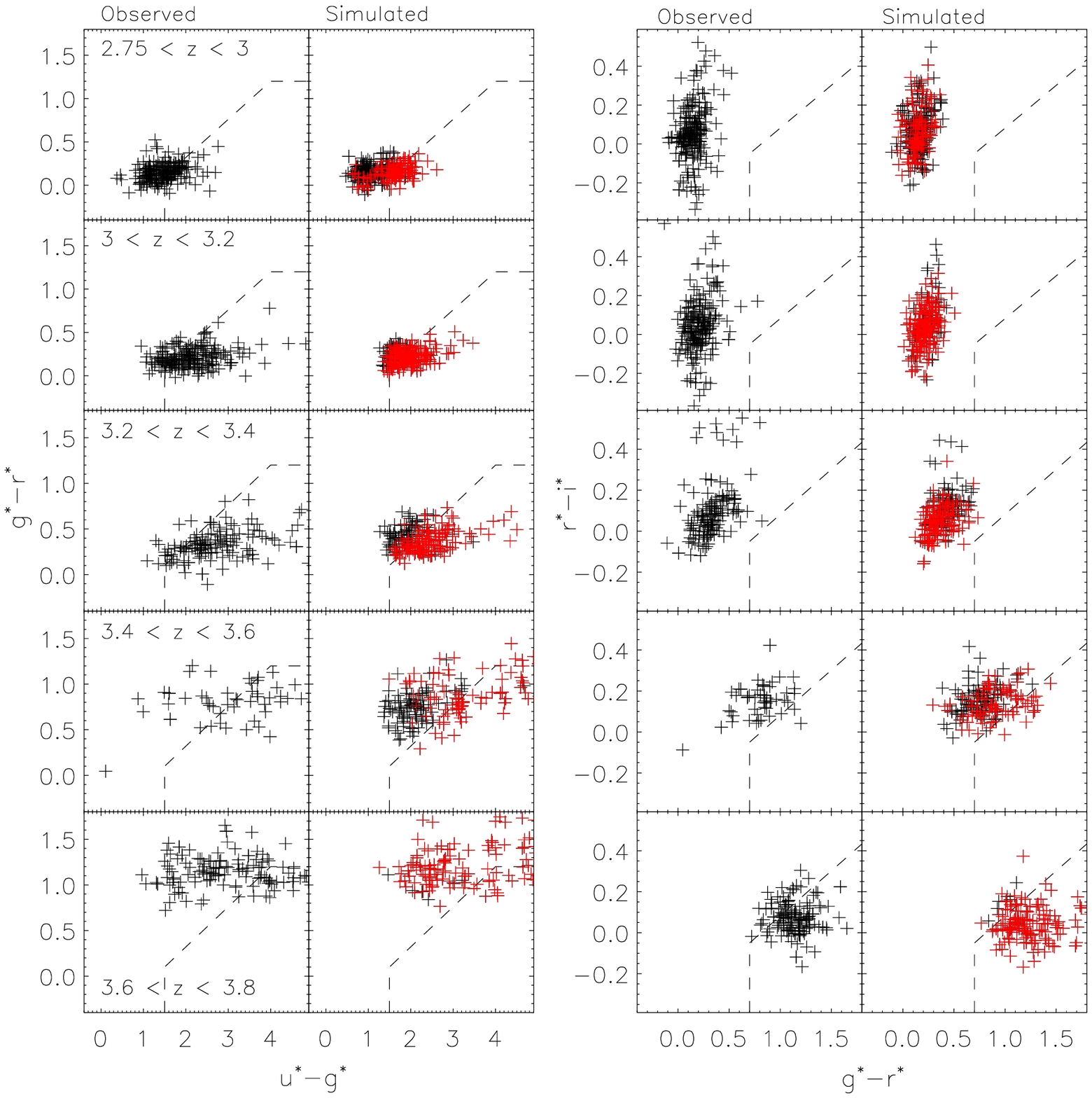}
\caption{Observed and simulated QSO colors as a function of redshift.  
Dashed lines show the color cuts used by the SDSS collaboration to 
identify QSO candidates; QSOs on the other side of the lines can also 
be selected if they meet the stellar locus outlier requirements.  
In the panels showing simulated colors, 
fainter symbols show objects which would have been selected, and darker 
symbols show objects which would not; the selection algorithm is least 
complete in the range $3.2<z<3.6$.}
\label{scolors}
\end{figure}

By convolving this spectrum with the SDSS filter curves, we could, in 
principle, generate mock luminosities and hence, mock colors.  
In practice, there are two reasons why we cannot do this quite yet.  
First, the SDSS spectra cover only a finite range in wavelength.  
By shifting a higher redshift spectrum blueward to simulate one at 
lower redshift, it may be that we have no spectrum left from which to 
estimate the flux in the reddest band $z^*$.  
To remedy this, we randomly choose a low-redshift QSO from among those 
observed with $z_{\rm obs}<3.2$, shift its spectrum to $z_{\rm sim}$, 
correct its flux blueward of Ly$\alpha$ to the required mean 
$D_A^{\rm fit}(z)$, and add Gaussian noise with rms $\sim 0.05$ if 
desired.  

\begin{figure}[t]
\plotone{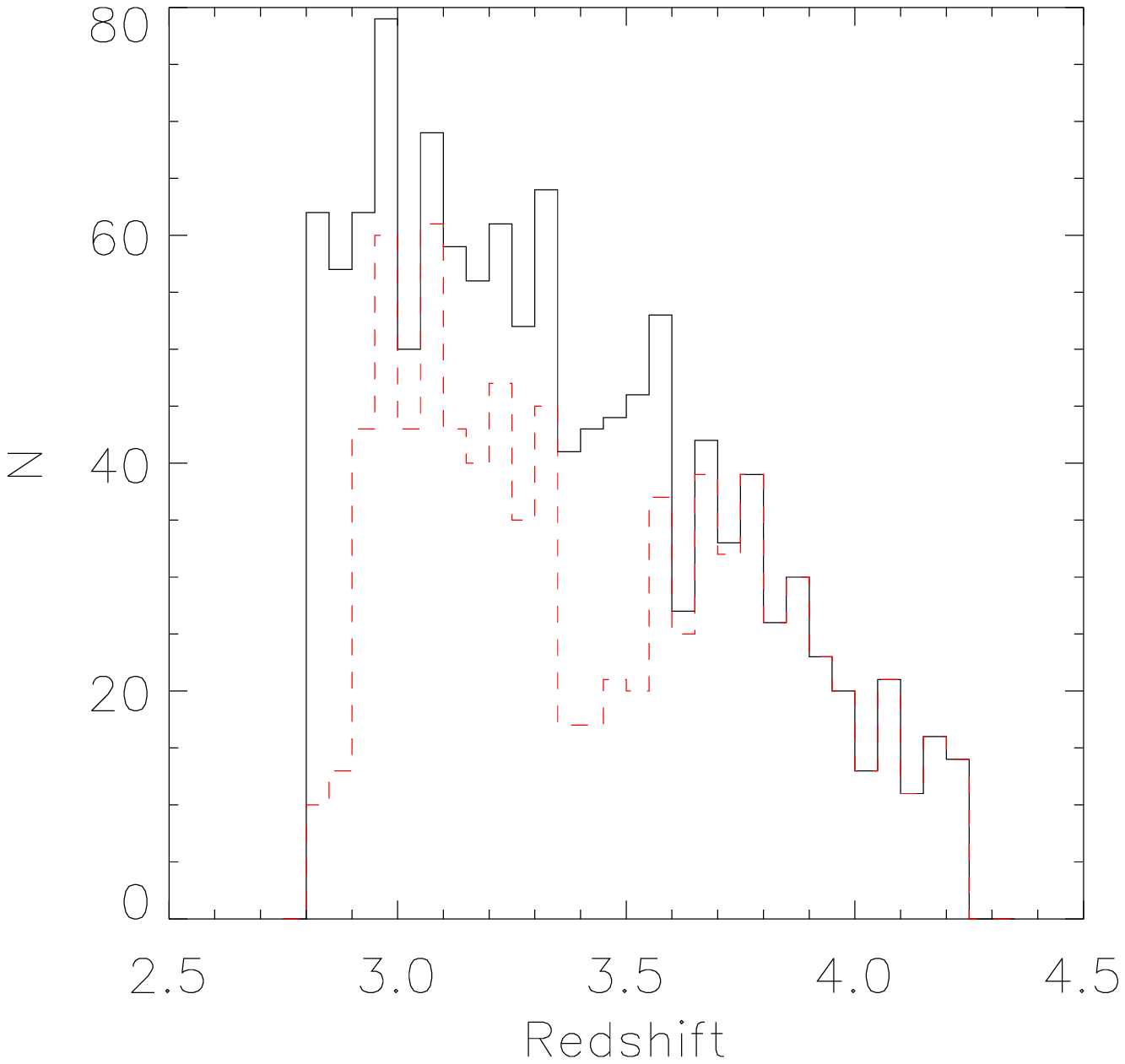}
\vspace{-4cm}
\caption{$N(z)$ distributions in the mock complete and SDSS--selected 
subsamples (solid and dashed histograms, respectively).  
Notice the selection effect at $z\sim 3.5$, which is similar 
to that seen in the observations (compare Figure~\ref{fig:NzQSO}).  }
\label{fig:Nzsim}
\end{figure}

\begin{figure}
\plotone{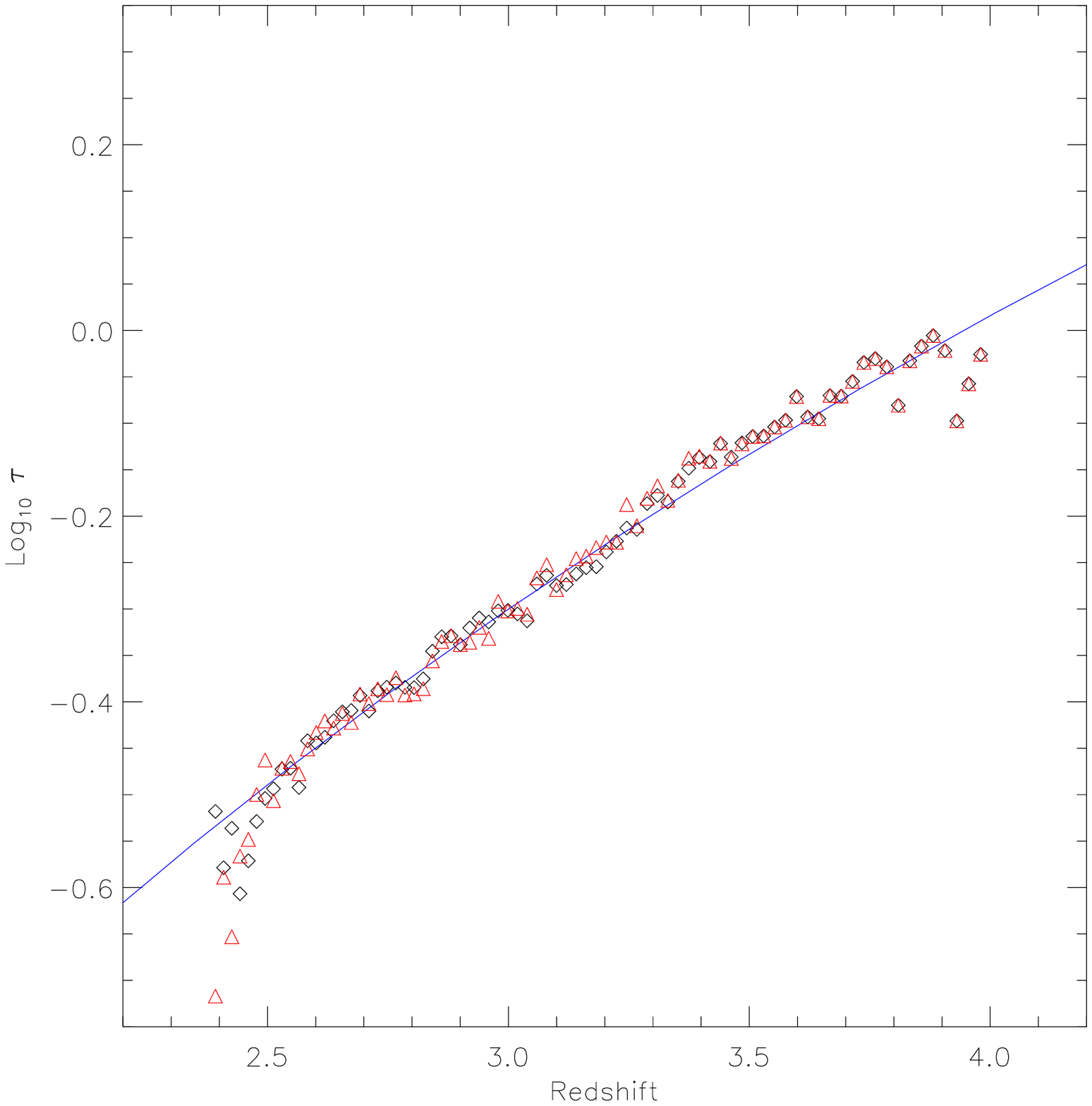}
\caption{Evolution of $\tau_{\rm eff}(z)$ in the simulated complete 
(diamonds) and SDSS--selected subsample (triangles).  
The complete sample follows the smooth input evolution (solid curve) 
as it should.  The evolution in the subsample is very similar.  
Notice in particular that the subsample does not show a feature at 
$z_\alpha \sim 3.2$, even though this is the regime in which the 
selection effects are strongest.  }
\label{tausim}
\end{figure}

Although this spectrum can be used to estimate the $z^*$ band flux, it 
may not cover the full wavelength range spanned by the bluest filter, 
$u^*$.  By averaging the two shifted spectra (normalized as described 
in Section~3 of the main text) in the range in which they overlap, 
and by simply using one spectrum in the regime where the other does not 
extend, we have a final simulated spectrum which spans the full required 
range in wavelengths.  (In practice, it sometimes happens that we still 
do not have enough wavelength coverage for the $u^*$ band.  This does 
not happen often, but when it does, we transform the spectrum of a 
$z_{\rm obs}>4.8$ QSO as described above, and we only add that piece of 
the spectrum which is needed to compute the $u^*$ band flux.)  

We can now perform the required convolutions with the five SDSS filter 
response curves, and so generate mock magnitudes for the five different 
bands.  Since these magnitudes are estimated from the spectra, we will 
refer to them as `spec--magnitudes'.  Before computing mock `spec--colors' 
using these mock spec--magnitudes, we must `flux--calibrate':  
we must check if the spec--magnitudes of the observed sample do indeed 
match the measured apparent magnitudes output by the SDSS photometric 
pipeline.  The finite length of the SDSS spectra means that we can only 
compute spec-magnitudes for the $g^*$, $r^*$ and $i^*$ bands, so this 
comparison can only be done for these three bands.  

The SDSS photometric pipeline outputs a variety of different measures 
of magnitude, two of which are useful for our purposes.  The first is 
the psf--magnitude, which is appropriate for point sources (see 
Stoughton et al. 2002 for details), and is the one used by the 
collaboration to define the colors which are used to determine whether 
or not an object is a QSO candidate.  The second, the fiber--magnitude, 
is an estimate of the light in a 3 arcsec aperture.  Since the spectra 
are taken with fibers of this size, spec--magnitudes are perhaps best 
compared with fiber--magnitudes.  

A comparison of the spec-- and fiber--magnitudes of the objects in our 
sample shows a linear relation with rms scatter around the mean of 
0.14~mags.  However, although a comparison of the psf-- and 
fiber--magnitudes of the objects shows the expected linear relation, 
there is a mean offset of 0.2~mags (the rms scatter around the mean 
is 0.07~mags).  This offset is approximately the same in all three bands.  
The spec-- versus psf--magnitude comparison also shows a mean offset:
$\langle {\rm fibermag - psfmag}\rangle\approx 0.2$~mags.  
Figure~\ref{smags} shows the difference between the spec--magnitudes 
measured from the spectra and the psf--magnitudes output by the SDSS 
photometric pipeline in the three bands, after this offset has been 
removed.  The plots show that the difference between the two does not 
correlate with psf--magnitude, and that the scatter between the two is 
$\sim 0.14$~mags.  
This shows that by subtracting 0.2~mags from the spec--magnitude one 
gets a reasonable estimate of the psf--magnitude.  
Thus, we can use the spec--magnitudes to compute spec--colors which 
are analogous to the psf--colors.  Since we can compute spec--colors 
from our simulated spectra, this allows us to compute mock $g^*-r^*$ 
and $r^*-i^*$ colors with which to model the effects of the SDSS 
selection algorithm.  

The SDSS QSO selection algorithm also makes use of the $u^*$ and $z^*$ 
band light.  We cannot compute spec-magnitudes in these bands from the 
observed spectra, but we can compare the psf-- and fiber--magnitudes 
in these bands.  They show the same offset as in the other three bands.  
So, although we cannot test if the spec-- and fiber--magnitudes are the 
same in these bands, it is likely that the spec-- and psf--magnitudes 
differ similarly to how the did in the other band.  
In principle, then, we could use the same procedure as for the other 
bands to convert from spec--magnitudes to mock $u^*$ and $z^*$ band 
psf--magnitudes, and hence to mock psf--colors.  

In practice, we have taken a different approach.  Namely, a plot of the 
observed $u^*-g^*$ psf--color versus the simulated $u^*_{spec}-g^*_{spec}$ 
spec--color shows a linear relation (with small scatter), but with an 
offset which we attribute to offset$_{u^*}$ (recall that we had previously 
calibrated and applied an offset to $g^*_{spec}$).  
An estimate of offset$_{z^*}$ is obtained analogously.  

Thus, we now have a mock QSO catalog with redshifts, luminosities in 
five bands, and hence colors.  The different panels in Figure~\ref{scolors} 
show the distribution of observed (left) and simulated (right) colors in 
different redshift bins.  The dashed lines show some of the SDSS selection 
cuts (the total set of selection criteria is described in 
Richards et al. 2002a).  In the panels which show the simulated colors, 
fainter symbols show objects which would have been targetted for 
observation by the SDSS collaboration, and darker symbols show objects 
which the collaboration would not have observed.  

On the whole, the simulated and observed samples are rather similar, 
although there are significant differences around $3.2<z<3.6$, and 
smaller differences at $z<3$.  
To quantify this, Figure~\ref{fig:Nzsim} compares the $N(z)$ distribution 
of the simulated complete sample (solid line) with the distribution 
of the subsample selected following the SDSS selected procedure (dashed 
line).  Notice how the simulated SDSS--subsample is missing objects in 
the range $3.2<z<3.6$.  Comparison with Figure~\ref{fig:NzQSO} shows that 
the resulting mock $N(z)$ distribution is rather similar to that seen 
in the data, suggesting that our mock catalogs are a reasonable model of 
the selection effect.  


Our concern is that this selection effect may be responsible for the 
feature we see in $\tau_{\rm eff}(z)$ (Figure~\ref{meantau}).  
To address this, Figure~\ref{tausim} compares $\tau_{\rm eff}(z)$ in the 
simulated complete sample with $\tau_{\rm eff}(z)$ in the SDSS--selected 
subsample.  The figure shows that the evolution of $\tau_{\rm eff}$ in 
both cases is very similar, even in the regime in which the SDSS--selected 
subsample contains many fewer objects than the complete sample.  
This suggests that the QSO selection does not give rise to the feature 
we see in Figure~\ref{meantau}.  

\subsection{The spectrograph}\label{app:Civ}
The feature in the Ly$\alpha$ forest at $z\sim 3.2$ occurs in the 
observed wavelength range $\lambda\sim 5000$~\AA.  
Our measurement makes strong demands on how well the SDSS spectrograph 
is calibrated.  The rest wavelength range 
$1430-1500$~\AA\ immediately blueward of the \ion{C}{4} emission line 
of most QSO spectra is relatively flat---indeed, as described in the 
main text, it is from within this region that we normalize the flux 
in each spectrum.  
We used this region to test whether inaccuracies in the spectrograph 
could have caused the feature we detected as follows.  

\begin{figure}
\plotone{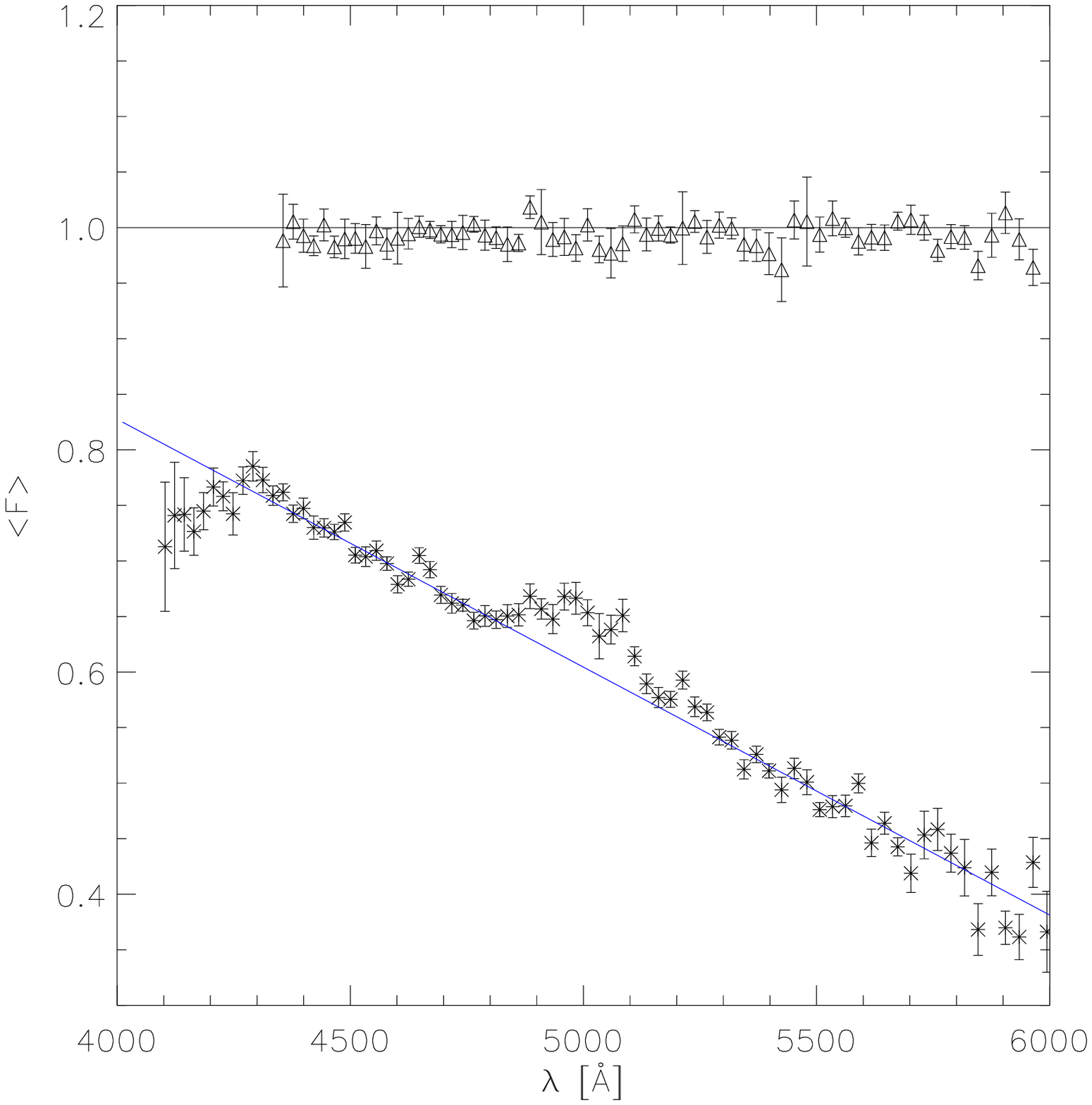}
\caption{The mean transmission in the Ly$\alpha$ forest 
(stars; rest wavelength range $1080-1160$~\AA) and the \Cfour\ forest 
(triangles; rest wavelength range $1430-1480$~\AA) of QSOs at 
$2 < z < 3$, as a function of observed wavelength.  
There is no feature at $\lambda_{obs}\sim 5000$~\AA\ 
in the mean \Cfour\ transmission (triangles), but there is a feature in 
the Ly$\alpha$ transmission (stars).}
\label{fig:tauciv}
\end{figure}

\begin{figure}
\plotone{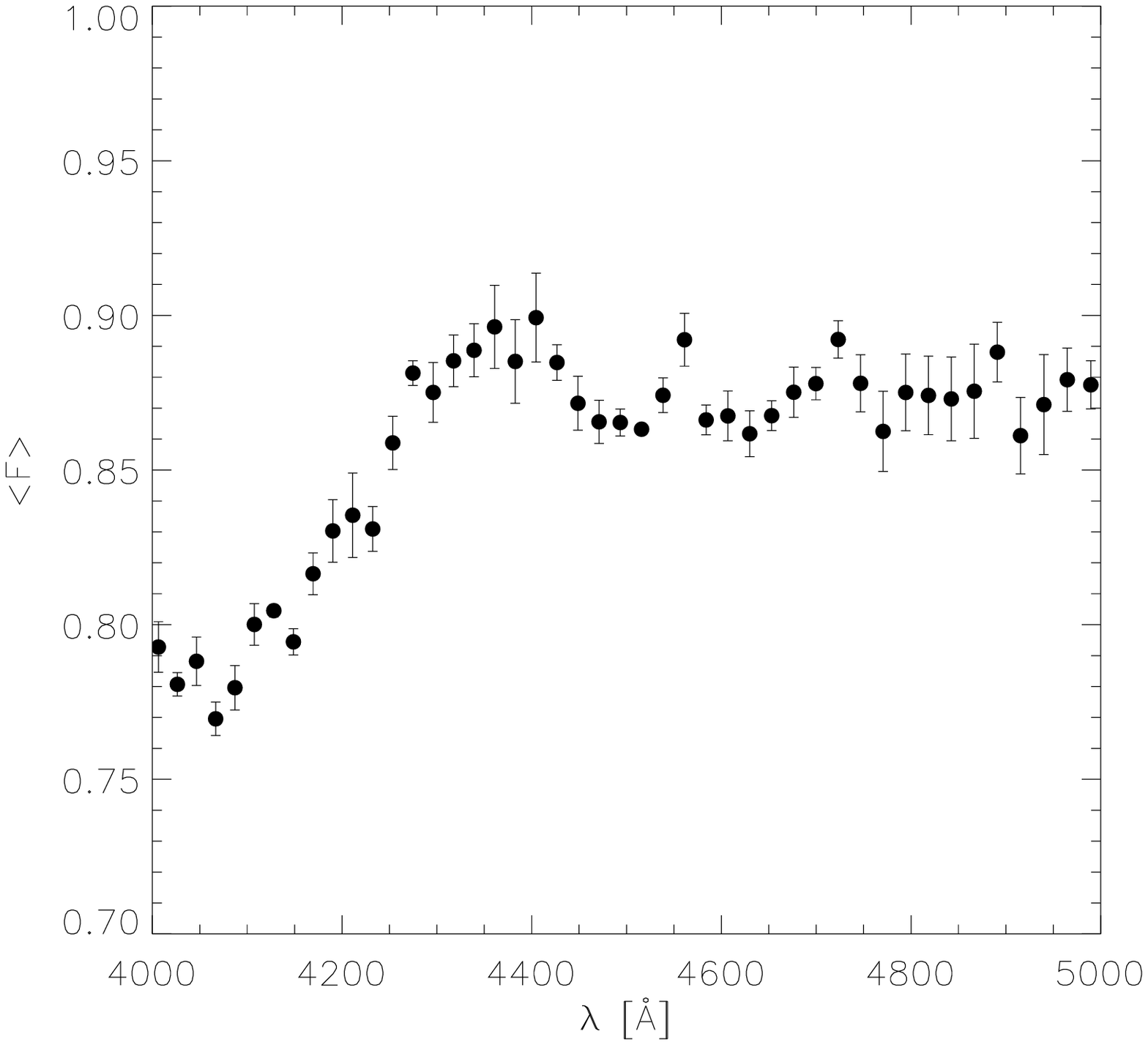}
\caption{Mean transmission in the \Cfour\ forest (rest wavelength 
range $1430-1480$~\AA) of QSOs at $1.75 < z < 2.5$, as a function of observed 
wavelength.  Spectra were normalized by the flux redward of \Cfour.  
There is a sudden drop in the observed flux at $\lambda_{obs}<4400$~\AA.  
The magnitude of the drop is similar to that seen in the Ly$\alpha$ forest 
of QSOs at $z\sim 2.8$, redshifted to the same observed wavelength.  
This similarity suggests that there is a problem with the spectroscopic 
calibrations:  $\lambda_{obs}\sim 4400$~\AA\ is close to the blue end of the 
spectrograph. }
\label{civcalib}
\end{figure}

Since the wavelength region blueward of the \Cfour\ emission line
lies $\sim 300$~\AA\ redward of the Ly$\alpha$ forest, to cover the same 
observed wavelength range it was necessary to use a sample of QSOs at 
lower redshift than the sample used in this paper.  Therefore, we 
extracted from the SDSS database 600 QSOs in the redshift interval 
$2 < z < 3$ (excluding BALs and/or low quality spectra).
We normalized each spectrum by the flux in the rest wavelength range 
$1350-1370$~\AA (the region blueward of \ion{Si}{4}).  
We then fit a continuum as described in the main text, 
and measured the mean transmission relative to this continuum in the 
wavelength range $1440-1480$~\AA, what we will call the \Cfour\ forest.  
Figure~\ref{fig:tauciv} compares the mean transmission versus observed 
wavelength in the Ly$\alpha$ forest (stars) and the \Cfour\ forest
(triangles).  Notice that there is no feature in the \Cfour\ forest.  
This indicates that calibration problems are not responsible for 
the feature we see in the Ly$\alpha$ forest.  

We stated in the main text that flux calibration problems near the 
blue end of the spectrograph limit the redshift range over which we 
can study the Ly$\alpha$ forest.  To illustrate the problem, we first 
selected 1000 QSOs at $1.75 < z < 2.5$.  We then calibrated each spectrum 
as follows.  Since \ion{C}{4} is at the blue end of the spectrum for 
the lower redshift QSOs, the region blueward of \ion{Si}{4} 
(the restframe wavelength $1350-1370$~\AA) is not measured, and so it 
cannot be used to calibrate.  The Vanden Berk et al. (2001) composite 
spectrum shows that the closest region redward of \ion{C}{4} which is not 
contaminated by emission lines is far away---at about 5000~\AA\ in the 
restrame!  On the other hand, although the region just redward of \ion{C}{4} 
is contaminated by emission, it is relatively featureless, and it does not 
appear to change very much in the redshift range $1.75 < z < 2.5$.  
Therefore, we used this region to calibrate the low redshift spectra, and 
we then measured the mean transmission in the \ion{C}{4} forest.  
Figure~\ref{civcalib} shows that the mean transmission is relatively 
constant redward of $\lambda_{obs}\sim 4400$~\AA; it is less than unity 
because the region just redward of \ion{C}{4} contains more flux than 
the region just blueward of \ion{C}{4} (cf. Figure~\ref{fig:composite}).  
However, the transmitted flux decreases by slightly more than ten percent 
between 4400~\AA\ and 4000~\AA.  

A drop of slightly more than ten percent is also seen in the Ly$\alpha$ 
forest at these same wavelengths.  There is no reason to believe that the 
Ly$\alpha$ optical depth at $z_\alpha\sim 4400/1215.67  - 1$ should be 
correlated with the \ion{C}{4} optical depth at $z\sim 2$, so the 
occurence of the feature almost certainly reflects a problem with the 
spectroscopic calibrations.  

\section{A skewed distribution for $\tau$}\label{app:skew}

\begin{figure}
\plotone{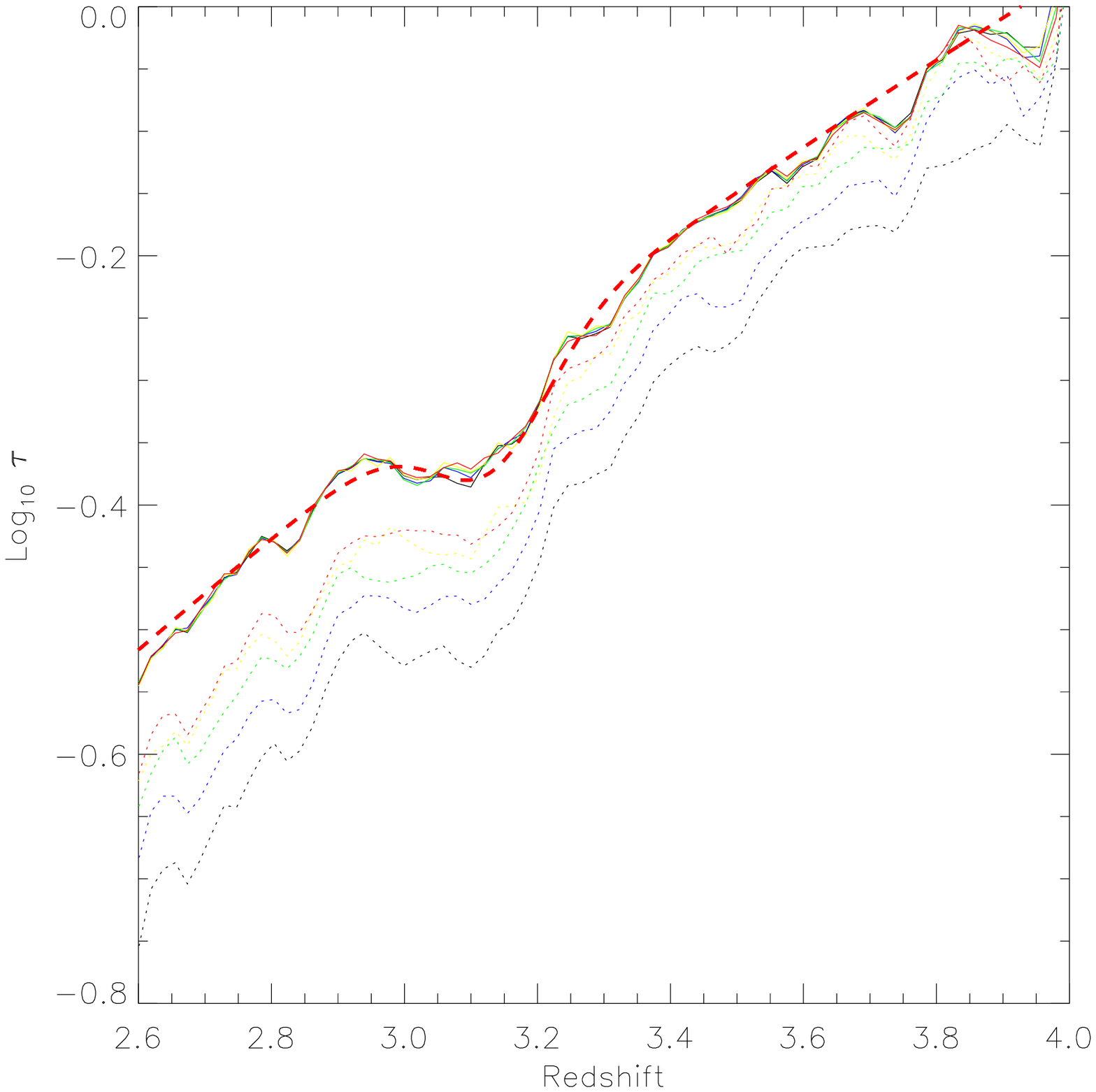}
\caption{Comparison of the mean (solid lines) and median (dotted lines) 
values of $\tau_{\rm eff}$ in the wavelength range $1080-1160$~\AA\ as 
a function of redshift. From bottom to top, dotted curves show 
$\tau_{\rm eff}$ computed splitting the Ly$\alpha$ forest using 
different bin sizes: 150~kms$^{-1}$, 300~kms$^{-1}$, 450~kms$^{-1}$, 
600~kms$^{-1}$, and 750~kms$^{-1}$.  
The mean values are insensitive to the bin size, whereas the median 
is strongly affected---this is a signature of the fact that the scatter 
around the mean is asymmetric.  }
\label{fig:MBtaumedian}
\end{figure}

In the main text we showed that our estimate of the mean transmission 
did not depend strongly on the length of the segments over which 
the measurement was averaged, provided this length was small.  Although 
the bin size is not important for mean statistics, the bin size does 
matter for median statistics.  

The solid curves in Figure~\ref{fig:MBtaumedian} show the evolution 
of $\tau_{\rm eff}$ computed using $\langle \bar F \rangle$,
the mean value of $\bar F$, as the bin size is increased from 
150~kms$^{-1}$ (bottom), to 300~kms$^{-1}$, 
450~kms$^{-1}$, 600~kms$^{-1}$, and 
750~kms$^{-1}$ (top).  The curves are indistinguishable from each 
other, illustrating that our results do not depend on the size of 
the bin.  
The dotted lines show the same, but for the evolution of $\tau_{\rm eff}$ 
computed using the median value of $\bar F$ rather than the mean.  
The Figure shows that, 
in this case, the bin size does matter --- the estimate from the median 
becomes increasingly similar to that from the mean as the bin size increases.  
Nevertheless, the $\tau_{\rm eff}$ does show a feature at $z\sim 3.2$, 
whatever the smoothing scale.  

That the median optical depth is smaller than the mean is a consequence 
of the well known fact that the distribution of flux decrements is skewed.  
Smoothing makes the fluxes in different pixels similar, so the median 
and mean values become increasingly alike as the smoothing scale is 
increased.

{}


\begin{thebibliography}{}

\bibitem[]{} Boksenberg, A. 1998, 
Structure and Evolution of the IGM from QSO Absorption Line Systems, 
eds. P. Petitjean \& S. Charlot (Paris: Ed. Fronti\`eres), 85
\bibitem[]{} Croft, R. A. C., Weinberg, D. H., Katz, N. \& Hernquist, L. 1997, 
ApJ, 488, 532
\bibitem[]{} Francis, P. J., Hewett, P. C., Foltz, C. B. \& Chaffee, F. H. 
1992, ApJ, 398, 476
\bibitem[]{} Frieman, J., et al. 2002, in preparation
\bibitem[]{} Fukugita, M., Ichikawa, T., Gunn, J. E., Doi, M., 
Shimasaku, K., \& Schneider, D. P. 1996, AJ, 111, 1748
\bibitem[]{} Gazta\~naga, E.~\& Croft, R. A. C., 1999, MNRAS, 309, 885
\bibitem[Gunn \& Peterson(1965)]{1965ApJ...142.1633G} Gunn, J.~E.~\& 
Peterson, B.~A.\ 1965, \apj, 142, 1633 
\bibitem[]{} Gunn, J. E. et al. 1998, AJ, 116, 3040
\bibitem[]{} Heap, S.~R., Williger, 
G.~M., Smette, A., Hubeny, I., Sahu, M.~S., Jenkins, E.~B., Tripp, T.~M., 
\& Winkler, J.~N.\ 2000, \apj, 534, 69 
\bibitem[]{} Hernquist, L., Katz, N.., Weinberg, D.H.,
Miralda-Escud\'e, J., 1996, ApJ, 457, L51
\bibitem[]{} Jenkins, E. B., Ostriker, J. P. 1991, ApJ, 376, 33
\bibitem[]{} Kim, T., Cristiani, S., \& D'Odorico, S. 2002, A\&A, 383, 747 
\bibitem[]{} Kriss, G.~A.~et al. 2001, Science, 293, 1112 
\bibitem[]{} Kulkarni, V. P., Huang, K.-L., Green, R. F., Bechtold, J., 
             Welty, D. E., \& York, D. G. 1996, MNRAS, 279, 197
\bibitem[]{} Lupton, R., Gunn, J. E., Ivezi\'c, Z., Knapp, G. R., 
\& Kent, S. 2001, in ASP Conf. Ser. 238, 
Astronomical Data Analysis Software and Systems X, 
ed. F. R. Harnden, Jr., F.~A.~Primini, and H. E. Payne 
(San Francisco: Astr. Spc. Pac.), p. 269 (astro-ph/0101420)
\bibitem[]{} Lynds, R. 1971, ApJL, 164, 73
\bibitem[]{} McDonald, P., Miralda-Escud{\' e}, J., Rauch, M., Sargent, W. L. W., Barlow, T. A., Cen, R., Ostriker, J. P. \ 2000, \apj, 543, 1 
\bibitem[]{} McDonald, P., Miralda-Escud{\' e}, J., Rauch, M., Sargent, W. L. W., Barlow, T. A., \& Cen, R. \ 2001, \apj, 562, 52 
\bibitem[]{} McDonald, P.~\& Miralda-Escud{\' e}, J.\ 2001, \apjl, 549, L11 
\bibitem[]{} Meiksin, A. 1994, ApJ, 431, 109
\bibitem[]{} Oke, J. B., \& Koryansky, D. G., 1982, ApJ, 255, 11
\bibitem[]{} Peebles, P. J. E. 1993, Principles of physical cosmology, 
Princeton University Press
\bibitem[]{} Press, W. H., Rybicki, G, B. \& Schneider, D. P. 1993, ApJ, 414, 64
\bibitem[]{} Rauch, M., Miralda-Escud{\' e}, J., Sargent, W. L. W., Barlow, T. A., Weinberg, D. H., Hernquist, L., Katz, N., Cen, R., \& Ostriker, J. P. 
1997, ApJ, 489, 7
\bibitem[]{} Rauch, M., 1998, ARA\&A, 36, 267
\bibitem[]{} Reimers, D., Kohler, S., Wisotzki, L., Groote, D.,
Rodeiguez-Pascual, P., \& Warmsteker, W., 1997, A\&A, 327, 890
\bibitem[]{} Richards, G. T., York, D. G., Yanny, B., Kollgaard, R. I., 
        Laurent-Muehleisen, S. A., \& Vanden Berk, D. E.  1999, \apj, 513, 576
\bibitem[]{} Richards, G. T., et al. 2002a, AJ, 123, 2945
\bibitem[]{} Richards, G. T., et al. 2002b, AJ, 124, 1
\bibitem[]{} Ricotti, M., Gnedin, N.~Y., \& Shull, J.~M.\ 2000, \apj, 534, 41
\bibitem[]{} Sargent, W. L. W., Steidel, C. C., \& Bocksenberg, A. 1989, ApJS, 69, 703
\bibitem[]{} Schaye, J., Theuns, T., Rauch, M., Efstathiou, G., \& 
Sargent, W.~L.~W.\ 2000, \mnras, 318, 817 
\bibitem[]{} Schneider, D. P., Schmidt, M., \& Gunn, J. E. 1991, AJ, 101, 2004 
\bibitem[]{} Songaila, A. \& Cowie, L. L. 1996, AJ, 112, 335
\bibitem[]{} Songaila, A. 1998, AJ, 115, 2184
\bibitem[]{} Songaila, A. \& Cowie, L. L. 2002, AJ, 123, 2183
\bibitem[]{} Steidel, C. C. \& Sargent, W. L. W. 1987, ApJ, 313, 171
\bibitem[]{} Stoughton, C. et al. 2002, AJ, 123, 485
\bibitem[]{} Theuns, T., Schaye, J., Zaroubi, S., Kim, T., Tzanavaris, P., \& Carswell, B. 2002a, ApJL, 567, 103
\bibitem[]{} Theuns, T., Zaroubi, S., Kim, T., Tzanavaris, P., \& Carswell, 
R. F. \ 2002b, \mnras, 332, 367
\bibitem[]{} Theuns, T., Bernardi, M., Frieman, J., Hewett, P., Schaye, J., 
Sheth, R. K., \& Subbarao, M. 2002, ApJL, 574, 111 
\bibitem[]{} Vanden Berk, D. E. et. al. 2001, AJ, 122, 549
\bibitem[]{} York, D. G. et al. 2000, AJ, 120, 1579
\bibitem[]{} Zaldarriaga, M., Hui, L., \& Tegmark, M. 2001, ApJ, 557, 519
\bibitem[]{} Zheng, W., Kriss, G. A., Telfer, R. C., Grimes, J. P., \& Davidsen, A. F. 1997, ApJ, 475, 469
\bibitem[]{} Zuo, L., \& Lu, L. 1993, ApJ, 418, 601

\end{thebibliography}
\end{document}